\documentclass[twocolumn]{aastex61}
\usepackage{amsmath}

\received{June 27, 2018}
\revised{\today}
\accepted{July 11, 2018}

\shorttitle{MSPs at low frequencies}
\shortauthors{Bhat et al.}

\def\psrone{{\rm PSR J0437$-$4715}}
\def\psrnameone{{\rm J0437$-$4715}}
\def\psrtwo{{\rm PSR J2145$-$0750}}
\def\psrnametwo{{\rm J2145$-$0750}}

\newcommand{\be}{\begin{eqnarray}}
\newcommand{\ee}{\end{eqnarray}}

\renewcommand{\vec}[1]{\mathbf{#1}}

\newcommand{\fnu}{\mbox{${f_{\nu} }$\,}}
\newcommand{\ft}{\mbox{${f _{\rm t} }$\,}}
\newcommand{\ftsq}{\mbox{${f ^2 _{\rm t} }$\,}}
\newcommand{\fti}{\mbox{${f _{\rm t,i} }$\,}}
\newcommand{\ftisq}{\mbox{${f ^2 _{\rm t,i} }$\,}}

\def\la{\mbox{\raisebox{-0.1ex}{$\scriptscriptstyle \stackrel{<}{\sim}$\,}}}
\def\ga{\mbox{\raisebox{-0.1ex}{$\scriptscriptstyle \stackrel{>}{\sim}$\,}}}

\newcommand{\Viss}{\mbox{${V _{\rm iss} }$\,}}

\newcommand{\Vmu}{\mbox{${V _{\rm \mu} }$\,}}

\newcommand{\Vbin}{\mbox{${V _{\rm bin} }$\,}}

\newcommand{\Vscreen}{\mbox{${V _{\rm scr} }$\,}}
\newcommand{\vecVeff}{\mbox{${\vec{V} _{\rm eff} }$\,}}
\newcommand{\Veff}{\mbox{${V _{\rm eff} }$\,}}

\newcommand{\Vobs}{\mbox{${V _{\rm earth} }$\,}}

\newcommand{\Veffsq}{\mbox{${V ^2 _{\rm eff } }$\,}}

\newcommand{\vecVeffperp}{\mbox{${\vec{V} _{\rm eff \perp} }$\,}}

\newcommand{\Parc}{\mbox{${P _{\rm arc} }$\,}}

\newcommand{\Tsys}{\mbox{${T _{\rm sys} }$\,}}
\newcommand{\Tant}{\mbox{${T _{\rm ant} }$\,}}
\newcommand{\Trcv}{\mbox{${T _{\rm rcv} }$\,}}

\newcommand{\Ae}{\mbox{${A _{\rm e} }$\,}}
\newcommand{\kB}{\mbox{${k _{\rm B} }$\,}}
\newcommand{\Apulse}{\mbox{${A _{\rm pulse} }$\,}}
\newcommand{\Speak}{\mbox{${S _{\rm peak} }$\,}}
\newcommand{\OmegaA}{\mbox{${\Omega _{\rm A} }$\,}}

\newcommand{\meanalpha}{\mbox{${ \langle \alpha \rangle }$\,}}

\newcommand{\pdist}{\mbox{${p _{\rm d} }$\,}}
\newcommand{\sd}{\mbox{${s _{\rm d} }$\,}}

\newcommand{\nud}{\mbox{${\nu _{\rm d} }$\,}}

\newcommand{\nudc}{\mbox{${\nu _{\rm d _{c} } }$\,}}
\newcommand{\dtnu}{\mbox{${d t / d \nu }$\,}}
\newcommand{\tauiss}{\mbox{${\tau _{\rm iss} }$\,}}

\newcommand{\cnsqunits}{${\rm m ^{-20/3}}$\,}

\newcommand{\dmu}{${\rm pc \ cm ^ {-3}}$\,}
\newcommand{\velu}{${\rm km \ s ^ {-1}}$\,}

\newcommand{\thetadiff}{\mbox{${{\rm \theta _{diff}}}$\,}}

\newcommand{\thetaref}{\mbox{${{\rm \theta _{ref}}}$\,}}

\newcommand{\Dos}{\mbox{${D _{\rm os}}$\,}}
\newcommand{\Dps}{\mbox{${D _{\rm ps}}$\,}}

\newcommand{\Cnsq}{\mbox{${ \overline { C _{\rm n} ^2 } }$\,}}

\begin{document}


\title{Observations of Low-Frequency Radio Emission from Millisecond Pulsars and Multipath Propagation in the Interstellar Medium}


\newcommand{\Curtin}{International Centre for Radio Astronomy Research, Curtin University, Bentley, WA 6102, Australia}

\newcommand{\ASU}{School of Earth and Space Exploration, Arizona State University, Tempe, AZ 85287, USA}

\newcommand{\USydney}{Sydney Institute for Astronomy, School of Physics, The University of Sydney, NSW 2006, Australia}

\newcommand{\CAASTRO}{ARC Centre of Excellence for All-sky Astrophysics (CAASTRO)}
\newcommand{\Ozgrav}{ARC Centre of Excellence for Gravitational-wave Discovery (OzGrav)}
\newcommand{\UToronto}{Dunlap Institute for Astronomy and Astrophysics, University of Toronto, ON, M5S 3H4, Canada}
\newcommand{\Victoria}{School of Chemical \& Physical Sciences, Victoria University of
	Wellington, Wellington 6140, New Zealand}
\newcommand{\UWM}{Department of Physics, University of Wisconsin--Milwaukee, Milwaukee, WI 53201, USA}
\newcommand{\UW}{Department of Physics, University of Washington, Seattle, WA 98195, USA}
\newcommand{\UWA}{International Centre for Radio Astronomy Research, University of Western Australia, Crawley 6009, Australia}
\newcommand{\CSIRO}{CSIRO Astronomy and Space Science, P.O. Box 76, Epping, NSW 1710, Australia}
\newcommand{\Swin}{Centre for Astrophysics and Supercomputing, Swinburne University of Technology, P.O. Box 218, Hawthorn, VIC 3122, Australia}
\newcommand{\Chalmers}{Department of Space, Earth and Environment, Chalmers University of Technology, Onsala Space Observatory, 439 92, Onsala, Sweden}
\newcommand{\Auckland}{Institute for Radio Astronomy and Space Research, Auckland University of Technology, 120 Mayoral Drive, Auckland 1010, New Zealand}

\correspondingauthor{N. D. R. Bhat}
\email{ramesh.bhat@curtin.edu.au}


\author[0000-0002-8383-5059]{N. D. R. Bhat}\affiliation{\Curtin} \affiliation{\CAASTRO}

\author[0000-0001-7662-2576]{S. E. Tremblay}\affiliation{\Curtin} \affiliation{\CAASTRO}

\author[0000-0001-6664-8668]{F. Kirsten}\affiliation{\Chalmers}

\author[0000-0001-8845-1225]{B. W. Meyers}\affiliation{\Curtin}  \affiliation{\CAASTRO} \affiliation{\CSIRO}

\author[0000-0001-5772-338X]{M. Sokolowski}\affiliation{\Curtin} \affiliation{\CAASTRO}

\author[0000-0003-2519-7375]{W. van Straten}\affiliation{\Auckland}

\author[0000-0001-6114-7469]{S. J. McSweeney}\affiliation{\Curtin} \affiliation{\CAASTRO}

\author[0000-0002-6380-1425]{S. M. Ord}\affiliation{\CSIRO}

\author[0000-0002-7285-6348]{R. M. Shannon}\affiliation{\Swin} \affiliation{\Ozgrav} \affiliation{\Curtin} \affiliation{\CSIRO} 


\author{A.~Beardsley}
\affiliation{\ASU}

\author{B.~Crosse}
\affiliation{\Curtin}

\author[0000-0002-4058-1837]{D.~Emrich}
\affiliation{\Curtin}

\author{T.~M.~O.~Franzen}
\affiliation{\Curtin}


\author{L.~Horsley}
\affiliation{\Curtin}

\author[0000-0003-2756-8301]{M.~Johnston-Hollitt}
\affiliation{\Curtin}

\author[0000-0001-6295-2881]{D.~L.~Kaplan}
\affiliation{\UWM}

\author{D.~Kenney}
\affiliation{\Curtin}

\author{M.~F.~Morales}
\affiliation{\UW}

\author{D.~Pallot}
\affiliation{\UWA}

\author{K.~Steele}
\affiliation{\Curtin}

\author[0000-0002-8195-7562]{S.~J.~Tingay}
\affiliation{\Curtin}
\affiliation{\CAASTRO}

\author{C.~M.~Trott}
\affiliation{\Curtin}
\affiliation{\CAASTRO}

\author{M.~Walker}
\affiliation{\Curtin}

\author{R.~B.~Wayth}
\affiliation{\Curtin}
\affiliation{\CAASTRO}

\author{A.~Williams}
\affiliation{\Curtin}

\author{C.~Wu}
\affiliation{\UWA}





\begin{abstract}	
Studying the gravitational-wave sky with pulsar timing arrays (PTAs) is a key science goal for the Square Kilometre Array (SKA) and its pathfinder telescopes. With current PTAs reaching sub-microsecond timing precision, making accurate measurements of interstellar propagation effects and mitigating them effectively has become increasingly important to realise PTA goals. As these effects are much stronger at longer wavelengths, low-frequency observations are most appealing for characterizing the interstellar medium (ISM) along the sight lines toward PTA pulsars.  The Murchison Widefield Array (MWA) and the Engineering Development Array (EDA), which utilizes  MWA technologies, present promising opportunities for undertaking such studies, particularly for PTA pulsars located in the southern sky.  Such pulsars are also the prime targets for PTA efforts planned with the South African MeerKAT, and eventually with the SKA. In this paper we report on  observations of two bright southern millisecond pulsars PSRs \psrnameone\ and \psrnametwo\ made with these facilities; MWA observations sampling multiple frequencies across the 80--250 MHz frequency range, while the EDA providing direct-sampled baseband data to yield a large instantaneous usable bandwidth of $\sim$200 MHz. Using these exploratory observations, we investigate various aspects relating to pulsar emission and ISM properties, such as spectral evolution of the mean pulse shape, scintillation as a function of frequency, chromaticity  in interstellar dispersion, and flux density spectra at low frequencies. Systematic and regular monitoring  observations will help ascertain the role of low-frequency measurements in PTA experiments, while simultaneously providing a detailed characterization of the ISM toward the pulsars, which will be useful in devising optimal observing strategies for future PTA experiments. 
\end{abstract}




\section{Introduction} \label{sec:intro}

The  detection of nanohertz gravitational waves is the primary science goal for current and future pulsar timing arrays (PTAs; \citealt{ppta,nanograv,epta,meertime+2017,ska+2015}). Success in this area will extend the spectrum of gravitational-wave (GW) astronomy, opened up by recent LIGO detections of stellar mass black-hole and neutron-star merger events \citep{ligo2016,ligo2017}, to frequencies where binary supermassive black holes are expected to dominate \citep{sazhin1978,detweiler1979}. 
With PTAs around the world achieving sub-microsecond or better timing precision, it has become important to carefully assess various contributing factors to the noise budget in timing data. These include the jitter noise arising from intrinsic pulsar emission processes that give rise to temporal variations of pulse shape or structure \citep[e.g.][]{ovh+11,shannon+2014,lam+2016}, delays and distortions to pulsar signals caused by interstellar propagation effects \citep[e.g.][]{cs2010,levin+2016}, and possible bias or uncertainties arising from instrumental effects including calibration and data processing \citep[e.g.][]{vanstraten2013,foster+2015}. Of these, the investigation of interstellar medium (ISM) effects has become a subject of considerable attention over recent years, since they may give rise to perturbations in timing measurements on timescales of $\sim$months to years, which can potentially result in low-frequency noise that can either mask, or subdue, the GW signatures in PTA measurements. 

The magnitudes of ISM effects are strong functions of the observing  frequency, $f$. For example, the dispersion delay due to propagation through cold plasma scales as $f^{-2}$, whereas pulse broadening due to multipath scattering inherent to propagation through turbulent plasma scales as $f^{-4}$ \citep[e.g.][]{bhat+2004}. Therefore, in the early days of PTA efforts, it was thought that ISM effects, particularly time-varying dispersion measure (DM), can be well modelled and corrected for using observational data that span well separated frequencies \citep[e.g.][]{ppta}. Further, it was also thought that effects of multipath scattering may not be a major concern for most PTA pulsars, given their low to moderate dispersion measures (DM \la 50 \dmu). However, recent investigations have significantly changed such a perception. For instance, from an observational perspective, applying DM corrections has proven  more difficult than previously thought  \citep[e.g.][]{keith+2013,lee+2014,jones+2017}. Moreover, phenomena such as extreme scattering events have been seen even in PTA data \citep[e.g.][]{coles+2015,kerr+2018}. 

There has also been a surge of theoretical efforts aimed at understanding more subtle effects caused by the ISM; for example, chromatic dispersion that can arise due to the frequency dependence of ISM propagation paths or volumes sampled \citep{cordes+2016}. The work of \citet{lam+2015} highlighted the need to develop observing strategies that optimally sample temporal variations of dispersion. Such strategies may also be pulsar-dependent, since the nature of the ISM can vary significantly between different sight lines. Further, the advent of wide-band instrumentation stimulated the investigation of developing optimal methods and strategies that allow the determination of pulse arrival times by taking into account the frequency evolution of pulse profiles over large bandwidths. This can also potentially enable more effective DM corrections \citep{pennucci+2014,liu+2014,lentati+2017}. However, 
owing to chromatic dispersion, correcting high-frequency dispersive delays using low frequency observations may not be straightforward
\citep{cordes+2016,shannon+2017}. 

These recent studies have shown that applying ISM corrections to PTA measurements is more complex than previously thought. Even though ISM effects can, in principle, be alleviated by resorting to shorter-wavelength observations ($\ga$3 GHz; \citealp{ppta+2015}), this is currently not a feasible option for the majority of PTA pulsars, at least within the sensitivity limitations of current facilities. Moreover, current PTA limits are largely dictated by a small number of well-timed pulsars \citep{nanograv+2016,epta+2015,ppta+2015}, and therefore detailed observational investigations of ISM effects can prove to be highly instructive. Applying whatever possible corrections, or even eliminating or de-prioritizing targets for which effective ISM corrections prove difficult, may help converge on more optimal observing strategies for future PTAs.  A detailed characterization of the ISM along the sight lines toward PTA pulsars would serve as a logical first step in that direction, for which observations at low frequencies are highly appealing. 

\begin{figure*}
\epsscale{1.05}
\plottwo{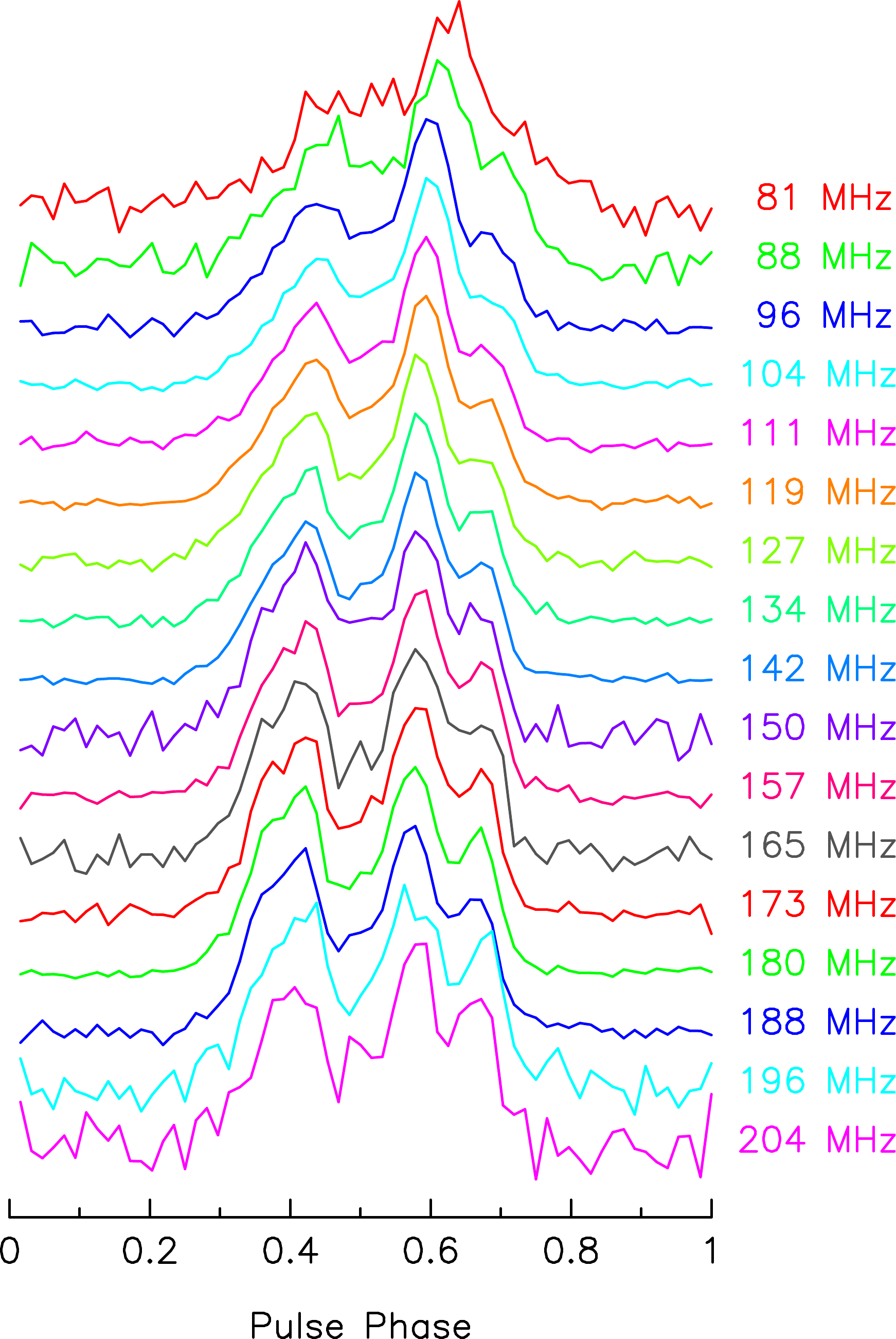}{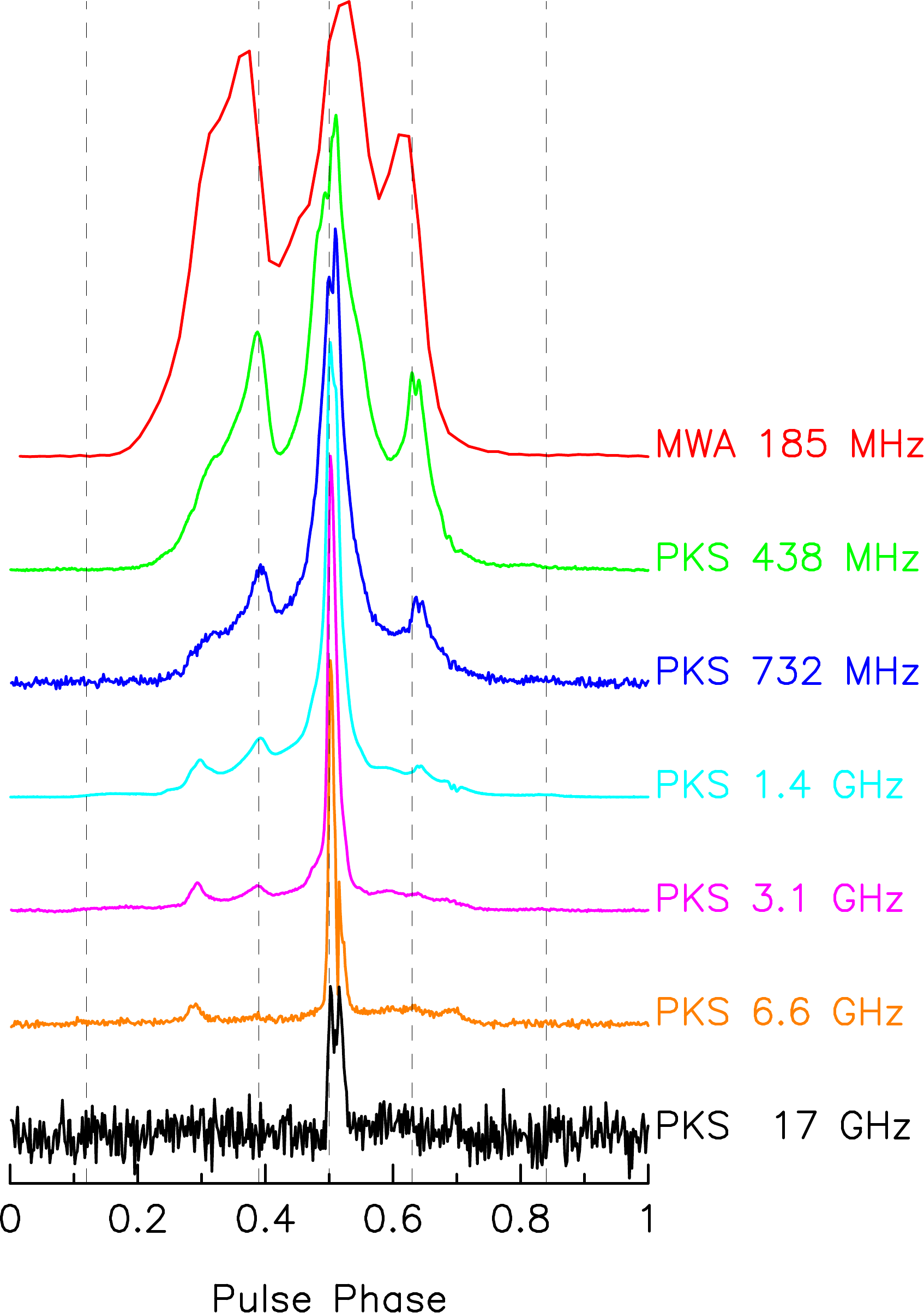}
\caption{
Integrated pulse profiles of \psrone\ at frequencies from 80 MHz to 17 GHz; {\it left}: MWA detections at multiple frequencies across the 80--205 MHz frequency range; {\it right}; MWA detection at 185 MHz along with Parkes detections (PKS) at higher frequencies. The detections shown on the left panel were made over fairly narrow bandwidths (1.28-MHz wide), whereas the 185-MHz detection ({\it right}) is from an observation over a 30.72-MHz bandwidth. MWA profiles are shown at a time resolution of 90 $\mu$s, with the residual dispersion smearing ranging from 25 $\mu$s (at 204 MHz) to 0.42 ms (at 81 MHz), whereas Parkes profiles have a much higher time resolution of 11-22 $\mu$s and are coherently de-dispersed. The dispersive smearing is only 45 $\mu$s for the MWA profile at 185 MHz (and hence negligible in comparison to the time resolution). 
\label{fig:f1}}
\end{figure*}

Over the past decade, a suite of new low-frequency facilities have become available for pulsar science: the Murchison Widefield Array (MWA) in Western Australia \citep{mwa2013}, the Long Wavelength Array (LWA) in the U.S.A. \citep{lwa2012},  and Low Frequency Array (LOFAR) in the Netherlands \citep{lofar2013}; all operating at frequencies below $\simeq$300 MHz. 
The MWA is also an official Precursor for the low-frequency component of the Square Kilometre Array (SKA-Low). 
Furthermore, the use of MWA Precursor technologies has led to the construction of the Engineering Development Array (EDA; \citealp{wayth+2017}) -- a 35-m SKA-Low station equivalent built out of MWA dipole elements.  Both are located in the shire of Murchison, where SKA-Low will be built. Therefore, notwithstanding their current limitations in achievable sensitivities, the MWA and EDA offer promising avenues for exploratory ISM studies of southern PTA pulsars, which will also be the prime targets for PTAs planned with the South African MeerKAT \citep{meertime+2017}, and eventually with the SKA \citep{ska+2015}. The MWA has also been upgraded with the addition of 128 new tiles (Phase II), and extending maximum baselines out to $\sim$6 km (Wayth et al. in prep.). The MWA and EDA allow pulsar observations at frequencies from 50 to 300 MHz, and can be exploited for a wide range of studies relating to pulsar emission physics and probing the ISM. 

In this paper, we report on some early observational science that leverages the large fractional bandwidths provided by 
the MWA and EDA. The focus of this paper is on the two brightest millisecond pulsars (MSPs) in the southern sky: \psrone\ and \psrtwo, which can be studied with both of these instruments. The combination of the MWA and EDA facilities has enabled their detections  down to $\sim$50 MHz. Observational data spanning such large fractional bandwidths are particularly useful for detailed low-frequency characterization, including profile and scintillation studies. In \S 2 we give a brief overview of the facilities used, with details on observations and data processing summarized in \S 3 and \S 4. Our main results are described in \S 5  and our further discussion and future work are summarised  in \S 6. Our conclusions are presented in \S 7. \\ 

\begin{figure}
\plotone{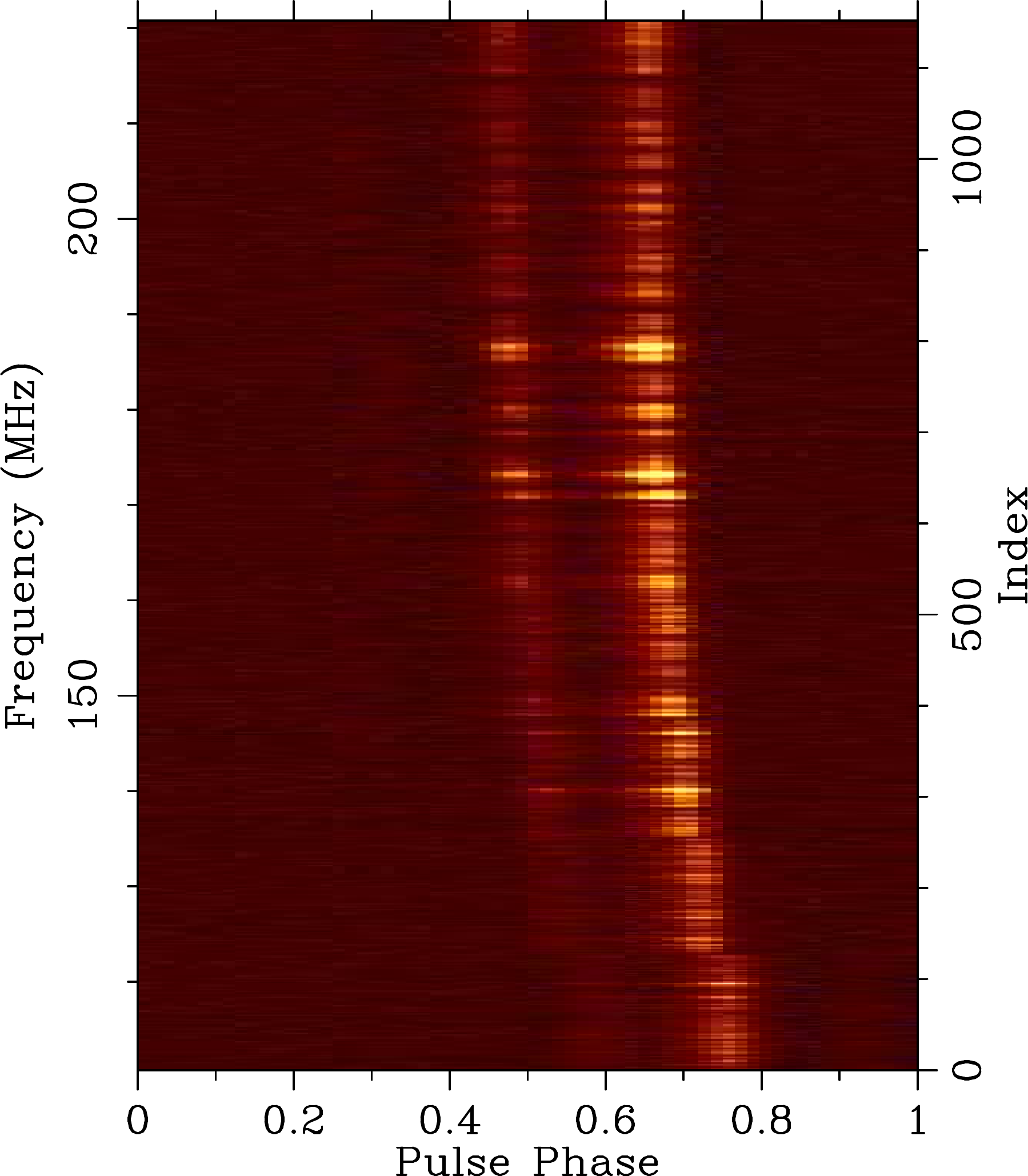}
\caption{
\psrtwo\ across the 110 to 220 MHz frequency range of the MWA, sampled in 9 sub-bands, each 2.56-MHz wide, at separations ranging from 8 to 16 MHz. The plot shows the pulse strength as a function of frequency vs. pulse phase, in which the 9 sub-bands are joined together. Data were processed using the catalog DM of 8.998 \dmu, and the quadratic sweep across the frequency range hence implies an excess DM of 0.006 \dmu. 
\label{fig:f2}}
\end{figure}


\section{Facilities} \label{sec:facilities}


\subsection{The MWA} \label{sec:mwa}

The MWA, originally conceived primarily as an imaging telescope, has been adapted to function also as a pulsar-capable facility through the integration of a voltage capture system (VCS) that follows the second stage poly-phase filterbank \citep{tremblay+2015}. This voltage-capture functionality allows the recording of $24\times1.28$ MHz from 
128 {\it tiles} (where each tile is a $4\times4$ dipole array), in both polarizations, at the native 100-$\mu$s, 10-kHz resolutions. The data rates are however large, 7.78 ${\rm GB\,s^{-1}}$ or approximately 28 ${\rm TB\,hr^{-1}}$, and the system is capable of recording a maximum duration of  $\sim$90 minutes at a given time. Because the MWA's hybrid  correlator \citep{ord+2015} was designed to generate visibilities at a rate no faster than once in 500 ms, the VCS capability has become the primary mode of observing  pulsars and fast transients. The recorded data are transported to the Pawsey Supercomputing Centre, where they can be run through various post-processing pipelines.

Even though the maximum recordable bandwidth of the VCS is limited to 30.72 MHz,  the flexible design of the MWA receiver and the signal path \citep{prabu+2015} can be exploited to leverage the large fractional bandwidth provided by the MWA by distributing the maximum recordable  bandwidth into multiple smaller sub-bands (each 1.28 MHz wide) so as to sample multiple frequencies {\it simultaneously}; e.g. 24$\times$1.28 MHz channels at $\sim$10-MHz separations to cover the nominal 70-300 MHz operating frequency range. This allows simultaneous observations at multiple different frequencies across the MWA's band, which is useful for a range of science goals; e.g., exploring the spectral evolution of the pulsar emission and investigating the frequency dependence of ISM effects such as dispersion and scattering. 

\begin{figure*}
\epsscale{1.1}
1.1\plottwo{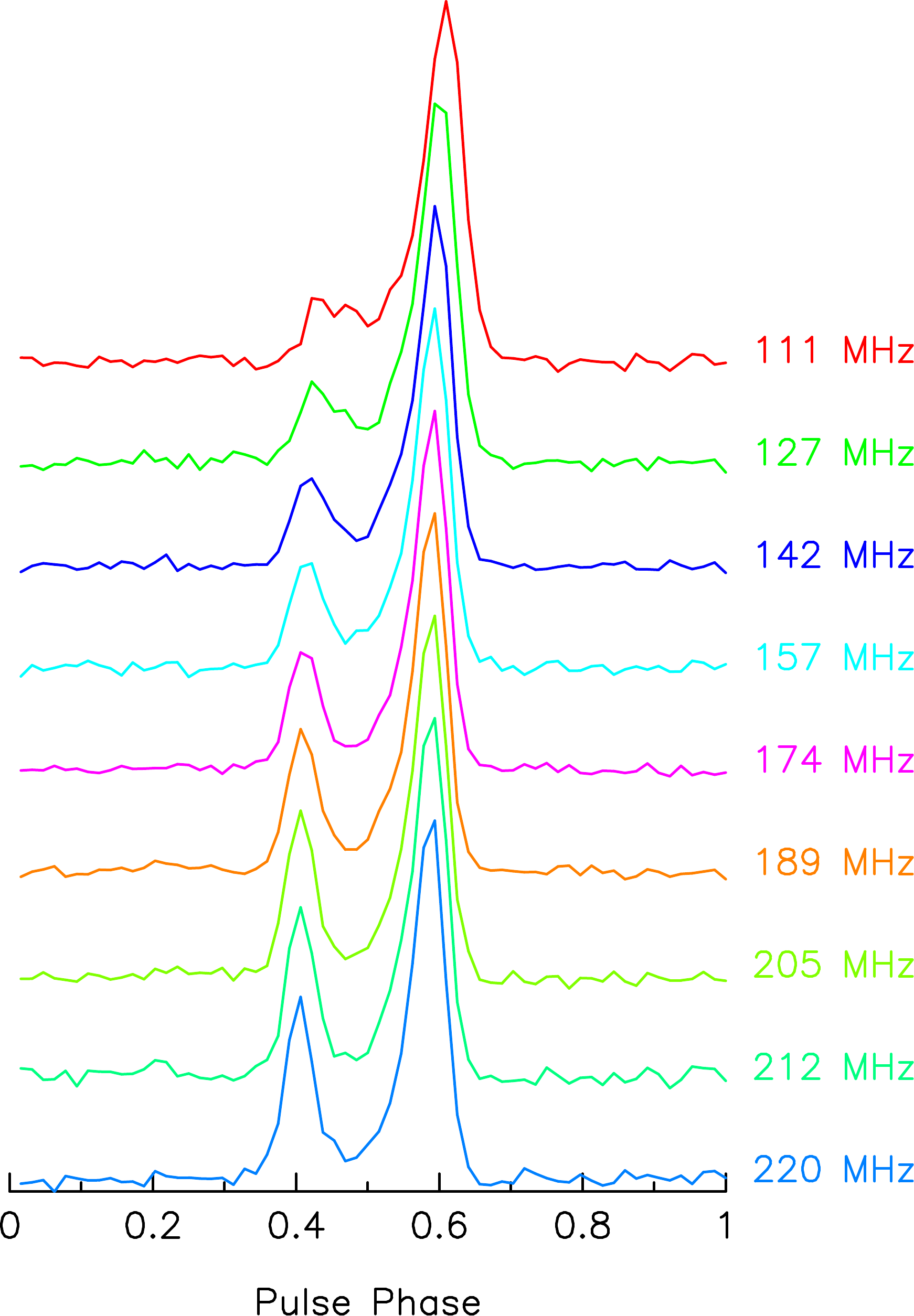}{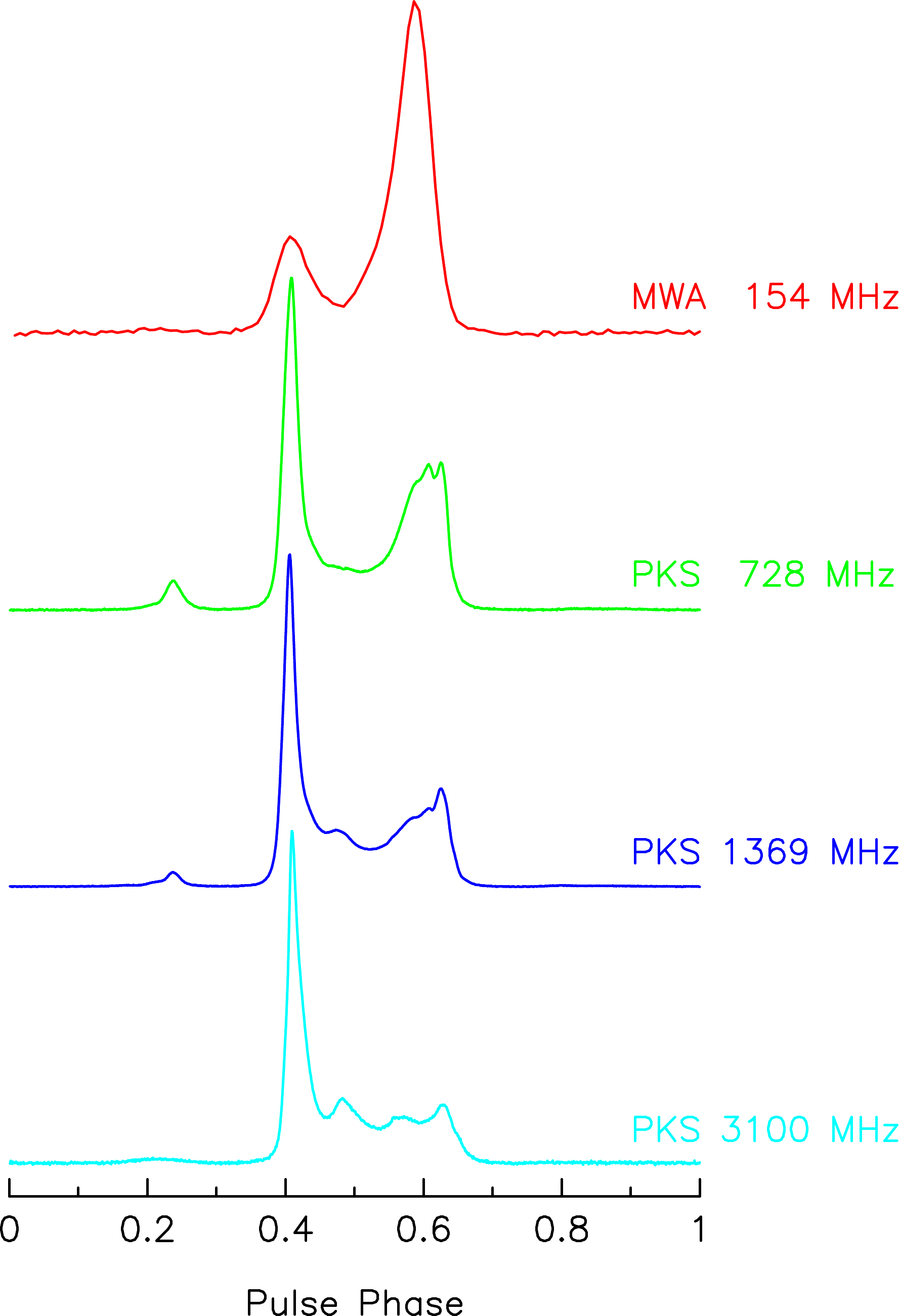}
\caption{
Integrated pulse profiles of \psrtwo\ at frequencies from 110 MHz to 3.1 GHz. {\it Left}: MWA detections at multiple frequencies across the 110--220 MHz frequency range; same data as shown in Fig.~\ref{fig:f2}, but after re-processing using the refined DM of 9.004 \dmu. {\it Right}: MWA detection at 154 MHz (over a contiguous frequency range of 140-170 MHz) along with Parkes detections at higher frequencies (data from \citealp{dai+2015}). The MWA profiles are at a time resolution of 125 $\mu$s, whereas the Parkes profiles are at a much higher time resolution of 8-16 $\mu$s 
(and are coherently de-dispersed). 
\label{fig:f2new}}
\end{figure*}

\subsubsection{The MWA upgrade} \label{sec:upgrade}

Observational data presented in this paper were obtained during 2016-2017 when the 
MWA underwent a major upgrade, which involved the deployment and commissioning of 
128 new tiles, effectively expanding on the original array, i.e., the Phase I MWA. Of the newly-added tiles, 72 are configured into 2 $\times$ 36 hexagonal layouts (each termed a ``Hex'') to provide a large number of redundant baselines, while the remaining 56 are placed further out from the original array to provide longer baselines up to $\sim$6 km (i.e. doubling of the imaging resolution and reducing the confusion limit). The basic system architecture, in terms of ability to correlate or  voltage-record is however limited to a maximum of 128 tiles at a given time. An important feature is that the array can be  periodically re-configured either as a {\it compact array}, in which the majority of the 128 tiles is located within a central region 300-metre across (comprising the two Hex's and the core),  or as an {\it extended array} that includes 56 outer tiles
and some fraction of the core. These new capabilities are formally referred to as the Phase II MWA. 

Even though the equivalent collecting area, or the achievable sensitivity, remains essentially the same as far as pulsar observations are concerned,  calibration and beam-forming with the compact configuration will naturally be far less susceptible to potential residual errors from ionospheric calibration.\footnote{In some rare cases, the achievable sensitivity may also be a function of the (synthesized) beam size, e.g., for observations of the Crab, the nebular emission is resolved with $\sim$5 km baselines of the extended Phase II array, resulting in increased sensitivity.} Such a compact configuration is also extremely appealing for undertaking large-area pulsar surveys given the significant reduction in the computational cost for realizing beam-forming across (or pixelising) the full field of view. 

\begin{figure}
\epsscale{1.161}
\epsscale{0.8}
\epsscale{1.163}
\plottwo{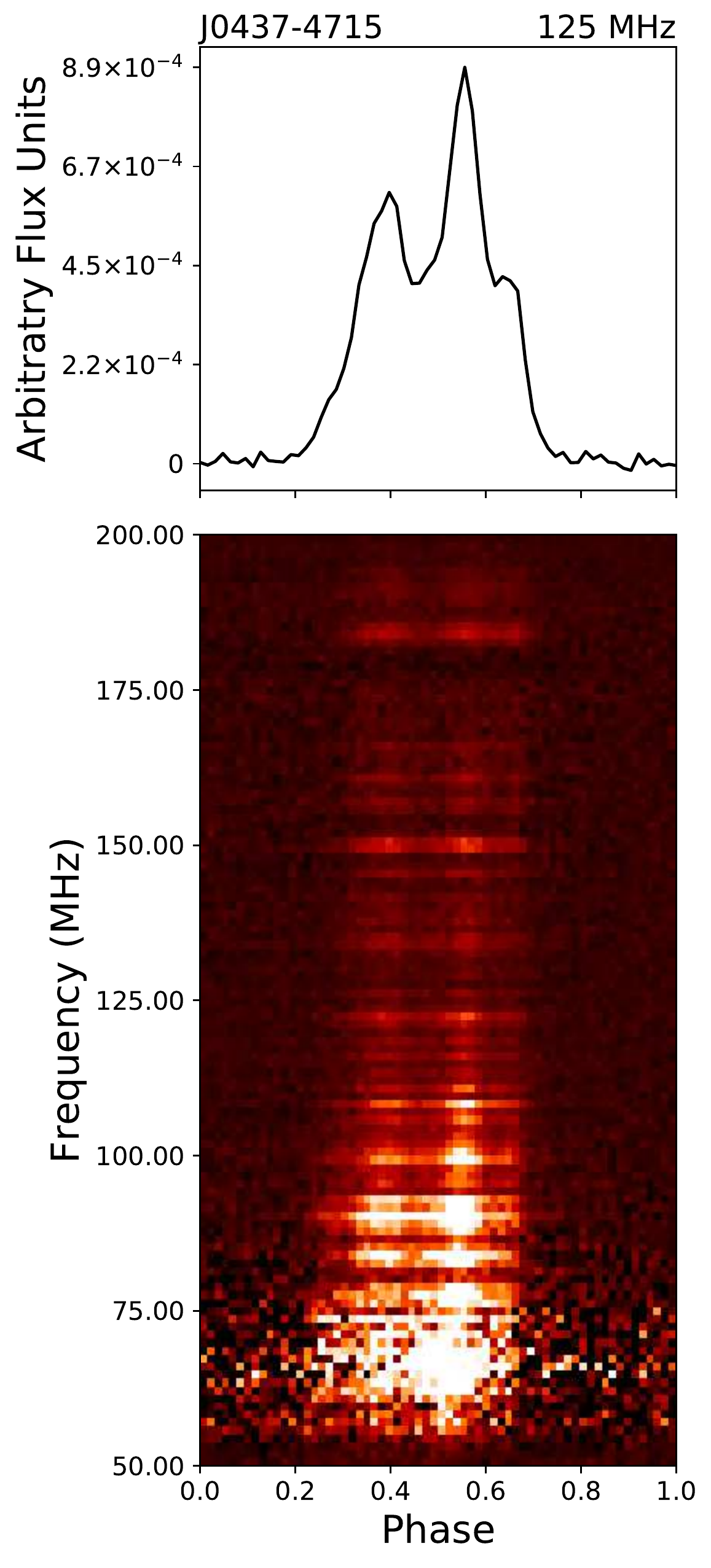}{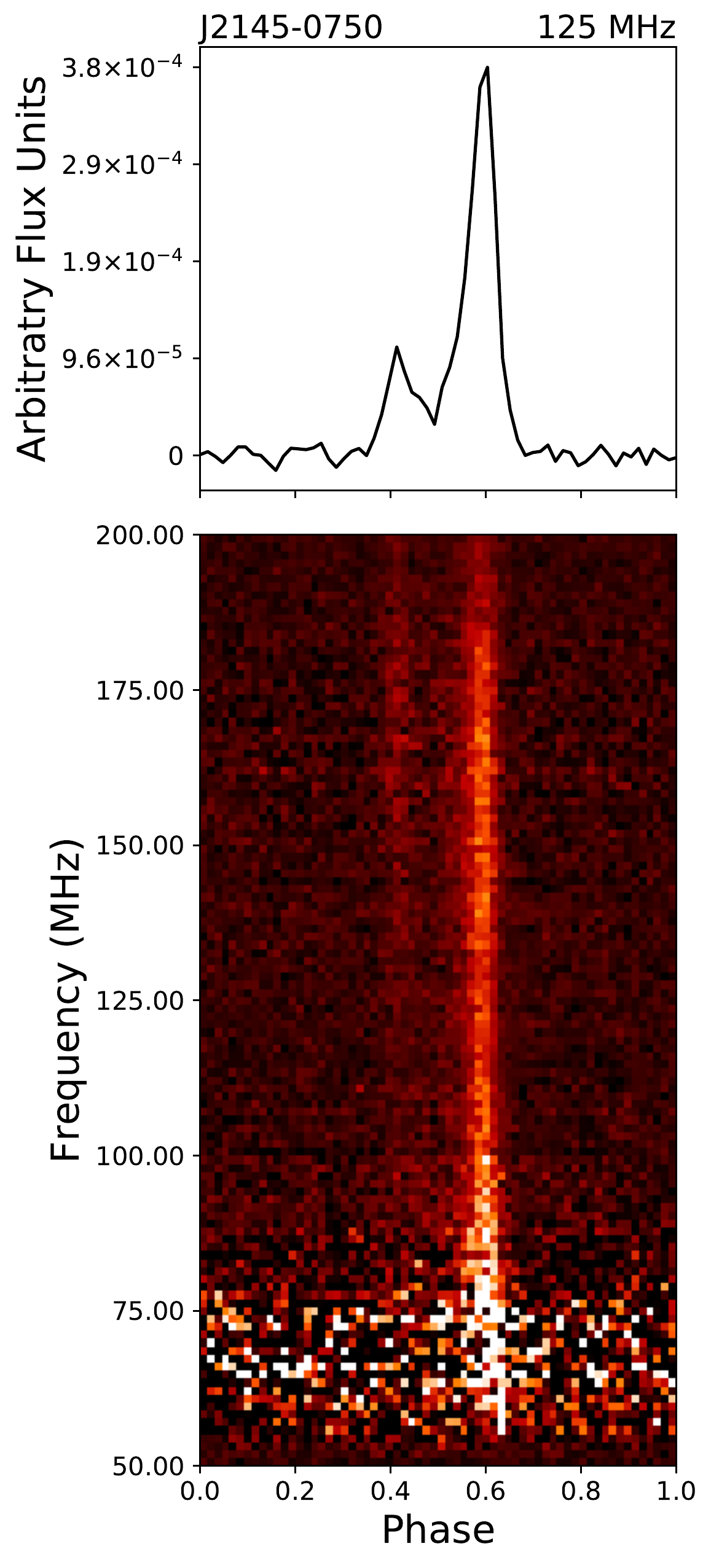}
\caption{
Pulsar detections across the EDA band: PSRs \psrnameone\ and \psrnametwo\ shown as time and frequency averaged pulse profiles (upper panels), and frequency vs. pulse phase (lower panels). Detections have been made down to 54 MHz for \psrone\ and 74 MHz for \psrtwo. Data were coherently de-dispersed using the convolving filterbank implementation within {\tt DSPSR} (see the text for details). \psrone\ data are from a single 5-minute observation, and are displayed at an effective time resolution of 90 $\mu$s, whereas for \psrtwo, multiple short observations (3-5 minutes each) over a 4-month time span were combined to produce an equivalent 2278-second observation, shown at an effective time resolution of 250 $\mu$s.
\label{fig:f3}}
\end{figure}


\subsection{The EDA} \label{sec:eda}

The EDA is effectively an implementation of the SKA station concept using MWA-style precursor technology. Details on its design, construction and commissioning are described in \citet{wayth+2017}.  It is an SKA-Low station equivalent built from 256 MWA dipoles configured to work as a phased array $\sim$35 m in size. The facility is located at the Murchison Radio-astronomy Observatory (MRO) and is hosted by the MWA under its external instrument policy.\footnote{http://www.mwatelescope.org/images/documents} The antenna elements are standard MWA dipoles with a low-noise amplifier design modified to extend the frequency coverage down to $\sim$50 MHz. A phased-array mode is realised through two stages of analog beamforming in the signal path; in the first stage, each group of 16 dipoles is connected to a standard MWA analogue beamformer, and hence 16 beamformers service the 256 dipoles that make up the array. The signals from these 16 beamformers are then connected via 200-m coaxial cables to a beamformer controller unit in a radio frequency (RF) shielded hut, where they are combined in the second stage beamformer, after appropriate delays, to make a phased array signal from the entire array. The data acquisition system then captures raw baseband data over the full 327.68 MHz band, and can record this on to disks. The pointing ability of the EDA is currently limited by the maximum delay possible in the MWA beamformer, and the array layout is such that each group of 16 dipoles make a sub-array of $\sim$10 m in size, which limits the maximum zenith angle (ZA) to 25$^{\circ}$ (i.e. a declination coverage from the equator to $\sim 50^{\circ}$ south). 

Even though primarily built for prototyping and verification purposes relating to SKA-Low development, we have been able to effectively turn the EDA into a pulsar-capable facility by integrating a pulsar processing pipeline to operate on the beam-formed voltage time series generated.
This allows observations to be made over a large instantaneous bandwidth of $\sim$50-300 MHz. Furthermore,  even with its limited pointing capability, pulsars such as PSRs \psrnameone\ and \psrnametwo\ can still be observed, albeit for fairly short durations of $\sim$5--10 minutes on a given day.\footnote{Longer observations are possible, albeit at the expense of reduced sensitivity, as a subset of dipoles will no longer be pointed toward the pulsar. In practice, the main limitation is the resources available for data acquisition and processing.} While this is indeed a limitation, the timing stability of the EDA has been pleasingly 
good and we have been able to synchronously combine data from multiple observations spanning up to several months, to yield good quality pulsar detections.

\section{Observations} \label{s:obs}

Observations of  PSRs \psrnameone\ and \psrnametwo\ were made with  the MWA and the EDA; the details  are summarized in Tables 1 and 2. Observations with the MWA made use of both the Phase I  and Phase II array configurations; the latter being in the compact configuration comprising the two Hex's and the core to provide a 128-tile array with a 300-m extent.   VCS recording was made in two 
different modes: i) over a contiguous band of 30.72-MHz bandwidth (24 $\times$ 1.28 MHz) to enable high-quality profile and scintillation studies,
and ii) by distributing the 30.72-MHz bandwidth into multiple smaller units and spacing them out to span a larger frequency range, thereby effectively realizing {\it simultaneous} multi-frequency observations across the MWA's 70-300 MHz frequency range. 
The latter allows, in effect, sampling a large frequency span, but at the expense of reduced sensitivity (per contiguous sub-band), and is particularly suited for studies of profile evolution across the MWA band and for high-precision DM determinations.  Both these capabilities are relatively new and mark significant  improvements over our early work \citep[e.g.][]{bhat+2014,bhat+2016}. For each pulsar, data were recorded over durations of 30 to 60 minutes, i.e. $\sim$14 to 28 TB per observation. Details on data processing procedures are summarized in \S~\ref{sec:data-mwa}.

In the case of the EDA, its limited pointing capability and the available computational resources for data processing  (for performing phase-coherent 
dedispersion over a large bandwidth) restricted our data recording to fairly short observations on a given day (typically $\sim$300 s).  The data rate is comparatively modest (4-bit samples, 2 polarizations, at the Nyquist rate of 655.36  ${\rm MB\,s^{-1}}$) and the typical observation results in $\sim$200 GB. Early test observations were made on 12 December 2016 (\psrone), and following the successful integration of the {\tt DSPSR} pipeline \citep{dspsr2011} for processing EDA data streams, further observations were made during April -- December 2017 (see Table 2 for details). Multiple observations were made of both \psrone\ and \psrtwo, and were combined later for improved detections. The details on data processing are briefly outlined in \S~\ref{sec:data-eda}.

\section{Data processing} \label{sec:data}

The high-time resolution data provided by the MWA and EDA signal paths are fundamentally different, and therefore require different processing pipelines for generating beam-formed time series at high time and frequency resolutions that are most amenable for pulsar detections. We briefly describe the basic steps in generating such time series data and the methods we employ for their flux calibration. Further details including calibration for full polarimetry will be detailed in an upcoming paper. 

\begin{figure}
\epsscale{0.8}
\epsscale{1.146}
\plotone{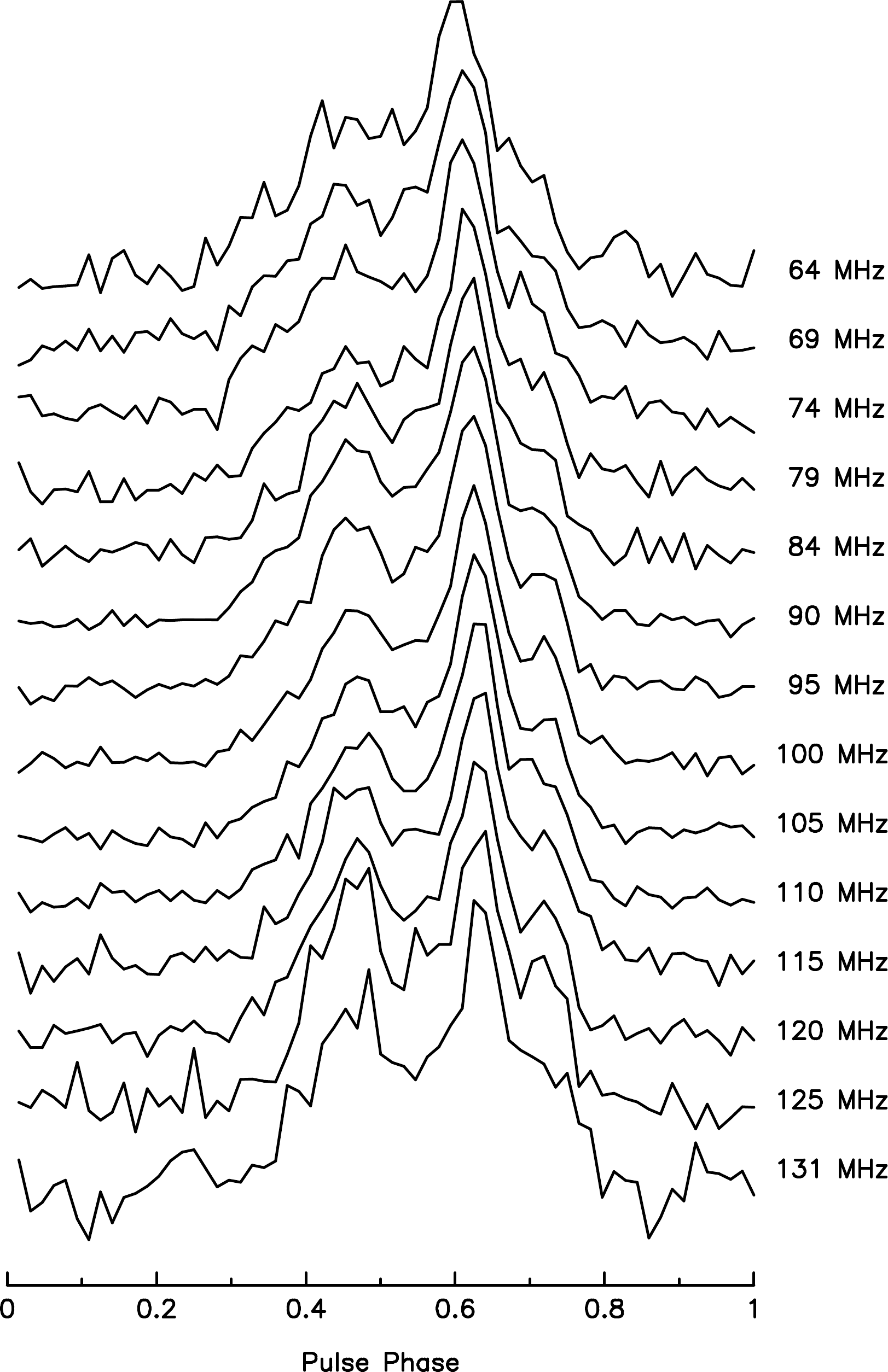}
\caption{
\psrone\ across the  64 to 131 MHz EDA band (i.e. a fractional bandwidth $\delta f /f $ = 0.5); at frequencies \la 131 MHz, the residual dispersive smearing is \ga 100 $\mu$s, which is the native resolution of MWA VCS. These profiles are from the same data as shown in Fig.~\ref{fig:f3}, i.e. a 5-minute observation on MJD=57743,  coherently de-dispersed while dividing the full EDA band into many sub-bands using a convolving filterbank (see the text for further details). Each profile is over a 5.12-MHz sub-band, at an effective time resolution of 90 $\mu$s, and can be compared to the MWA profiles at nearby frequencies (shown in Fig.~\ref{fig:f1}), where the DM smear ranges up to 0.42 ms. 
\label{fig:f3new}}
\end{figure}

\subsection{Generation of beam-formed time series} \label{sec:data-proc}

\subsubsection{The MWA} \label{sec:data-mwa}

VCS-recorded data can be coherently summed to generate a tied-array (phased array) beam on the sky by performing a coherent addition of voltage signals from individual tiles (Ord et al. in prep.). 
This maximizes the achievable sensitivity and is thus the preferred mode for pulsar observations with the MWA.  
However,  this  procedure involves incorporating polarimetric beam responses of individual tiles (using analytic models for the beams), as well as the cable and geometric delay models and complex gain information (amplitude and phase) for each tile. The calibration can make use of either an in-field bright source, or multiple moderately-bright sources in the field, or sometimes,  pointed observations of one of the bright calibrators  (e.g., Hydra A, 3C444, Hercules A, etc.) taken prior to pulsar observations. In all these cases,  first the visibilities are created using  the MWA correlator \citep{ord+2015}, and then for each sub-band, a calibration solution (for both amplitude and phase) is iteratively generated for each tile using the Real Time System (RTS; \citealp{mitchell+2008}). The output from the RTS is a calibration solution for each coarse channel, i.e. 24 solutions per observation for each tile. Between iterations,  tiles with poor quality calibrations solutions are flagged out (the number tends to vary depending on the observation). Once solutions have converged for each sub-band a coherent tied-array beam is produced using only those tiles with good solutions.  The full processing chain runs on the Galaxy cluster of the Pawsey supercomputing facility and is routinely employed for sensitive pulsar  observations 
\citep[e.g.][]{mcsweeney+2017,meyers+2017}.

There is however an important caveat associated with the choice of the in-field calibration strategy. Since the tile (primary) beam pattern is a strong function of the observing frequency, if the calibrator is not in close proximity to the pulsar, it may lead to difficulties calibrating higher-frequency bands as the calibrator becomes near (or outside) the edge of the beam (and hence not ideal for reliable calibration). For example, Pictor A, which is located at $\sim10^{\circ}$ from the position of \psrone, is no longer a suitable in-field calibrator for this pulsar at frequencies \ga 220 MHz. Observations across the full MWA band may therefore warrant considering additional (or alternate) strategies to ensure successful calibration  across the full frequency range. 

				
\begin{deluxetable*}{ccCccc}[t]
\tablecaption{Observational parameters with the MWA  \label{tab:mwa}}
\tablecolumns{6}
\tablenum{1}
\tablewidth{0pt}
\tablehead{
\colhead{PSR} &
\colhead{MJD of} &
\colhead{Array} &    
\colhead{Frequency range} & {VCS recording mode} & 
\colhead{Observing } \\
\colhead{} & \colhead{observation} & \colhead{configuration} & 
\colhead{(MHz)} & \colhead{Sub-bands (MHz)} & \colhead{duration (s)}
}
\startdata
\psrnameone & 57051 & {\rm Phase ~I} &   80-300 & 24$\times$1.28  & 1863     	\\
\psrnametwo & 57701 & {\rm Phase~II}$^a$ &   80-220$^b$ & 12$\times$2.56  & 1884	\\
\psrnametwo & 57704 & {\rm Phase ~II}$^a$ & 140-170 &       1$\times$30.72  & 2497	\\
\psrnameone & 57718 & {\rm Phase ~II}$^a$ & 170-200 &       1$\times$30.72  & 3605	\\
\enddata
\tablenotetext{a}{The compact configuration comprising 2 x Hex's and the core (see the text for details).}
\tablenotetext{b}{The frequency range was restricted to below 220 MHz following the calibration difficulties of initial observations.}
\end{deluxetable*}

\subsubsection{The EDA} \label{sec:data-eda}

The output of the second stage beamformer in the EDA signal path is low-pass filtered and sampled at a rate of 655.36 MHz. 
For pulsar observations, the resultant data streams are re-scaled to 4-bit samples before writing out to disks as a dual-polarization voltage time series. These data can therefore be processed using the phase-coherent dispersion removal technique over the majority of the 327.68 MHz band, enabling pulsar observations over a large instantaneous bandwidth.  
In practice, the available bandwidth is limited by the response range of MWA dipoles (i.e. $\sim$50-300 MHz) and the rapid growth of dispersive delay at low radio frequencies. The {\tt DSPSR} software package\footnote{http://dspsr.sourceforge.net}, which has been the backbone of multiple generations of pulsar backends at the Parkes 64-m telescope, has been integrated with the output of the EDA data stream.  
The baseband data stream can be coherently de-dispersed while dividing the full EDA band into sub-bands (typically 32k channels) using the convolving filterbank \citep{dspsr2011} to generate virtually artefact-free pulse profiles.
For a given DM, the lowest radio frequency amenable to coherent de-dispersion depends on the computational memory available for processing.

\subsection{Flux density calibration} \label{sec:flux}

The MWA and EDA are promising instruments for investigating the low-frequency regime of flux density spectra. However, such aperture array instruments require suitably devised approaches for calibrating pulsar flux densities. Below we describe the methods employed for the initial studies presented in this paper.

\subsubsection{MWA detections} \label{sec:mwaflux}

Flux calibration for MWA pulsar detections made use of the technique developed by \citet{meyers+2017}. In summary, we simulated the tied-array (phased-array) beam pattern by modeling it as the product of the  tile beam pattern and the array factor (see their equations 11 and 12). The tile beam pattern was simulated using the formalism as per \citet{sutinjo+2015} and the array factor depends on the configuration used for the observation (i.e. whether Phase I vs.  Phase II, and the number of tiles used to form a tied-array beam on the pulsar). The simulated tied-array beam is then convolved with the global sky model (GSM; \citealp{gsm+2008}), and integrated over the sky (see \citealp{marcin+2015}) to produce  estimates of the antenna temperature \Tant. These are then used to calculate the system temperature \Tsys, under the assumption that the radiation efficiency $ \eta _{\rm rad} \simeq 1 $ and using the receiver temperature measurements ($T_{\rm rcv} \approx 40$ K) for the MWA. The integral of this array factor power pattern is then used to determine the beam solid angle $\OmegaA$, from which the tied-array gain $G=\Ae/2\kB$ (where \Ae\ is the effective collecting area and \kB\ is the Boltzmann constant) is estimated and translated into the system equivalent flux density 
${\rm SEFD}$=\Tsys/$G$. 
Using this approach, we estimated $G$ and \Tsys for each sub-band of observations across the MWA band, by factoring in also the directional and frequency dependencies of the MWA tile pattern. 
We then use the standard radiometer equation for pulsar observations (i.e. accounting for the pulsar duty cycle $W/P$, where $W$ is the equivalent pulse width and $P$ is the pulsar period) to calculate the corresponding flux density scales for our observations \citep{handbook}. 
The equivalent width $W$ is estimated as $\Apulse/\Speak$, where \Apulse is the integrated flux within the ``on" pulse region and \Speak is the peak amplitude of the detected pulse. 
These are then used to estimate the flux density scales at multiple frequency bands of MWA observations.

\subsubsection{EDA detections} \label{sec:edaflux}

As described in \citet{wayth+2017}, the performance of the EDA is well characterized, particularly in terms of the receiver and system temperatures, the beam and sky models and the sensitivity.  Further, as shown in their Fig. 9, the predicted sensitivity based on the beam and sky models is found to be in good agreement with measurements. Therefore, it is justifiable to use the EDA beam model to calculate the effective collecting area $A_e$ and the system temperature \Tsys, and consequently, the sensitivity $A_e/\Tsys$, or  ${\rm SEFD} = 2 k_B A_e/\Tsys$, for the specific pointing directions of PSRs \psrnameone\ and \psrnametwo. The system temperature $\Tsys \cong \eta _{\rm rad} \Tant+\Trcv$, where \Trcv\ is the receiver temperature and \Tant\ is the antenna temperature, calculated as the beam-weighted average sky temperature, using the well-known sky model of \citet{haslam+1982} at 408 MHz (see eq. 3 in \citealt{wayth+2017}), scaled to the observing frequency as $T(f)=T_{408}(f_{\rm MHz}/408)^{\beta}$, where $\beta=-2.55$, $f_{\rm MHz}$ is the frequency (in MHz) and $T_{408}$ is the sky temperature at 408 MHz. 
As all our  EDA observations were made near the meridian transit, this provides fairly reliable estimates of the sensitivity.  
The values of $G$ and \Tsys\ obtained in this manner are then used to estimate the flux density scales at multiple frequency bands of EDA observations (i.e. in 5.12-MHz sub-bands, across the EDA band), in a manner similar to the one for MWA pulsar detections (see \S~\ref{sec:mwaflux})\footnote{We exclude the sub-bands in which no meaningful detections were made, or those severely corrupted by radio frequency interference.}


\begin{figure*}
\gridline{\fig{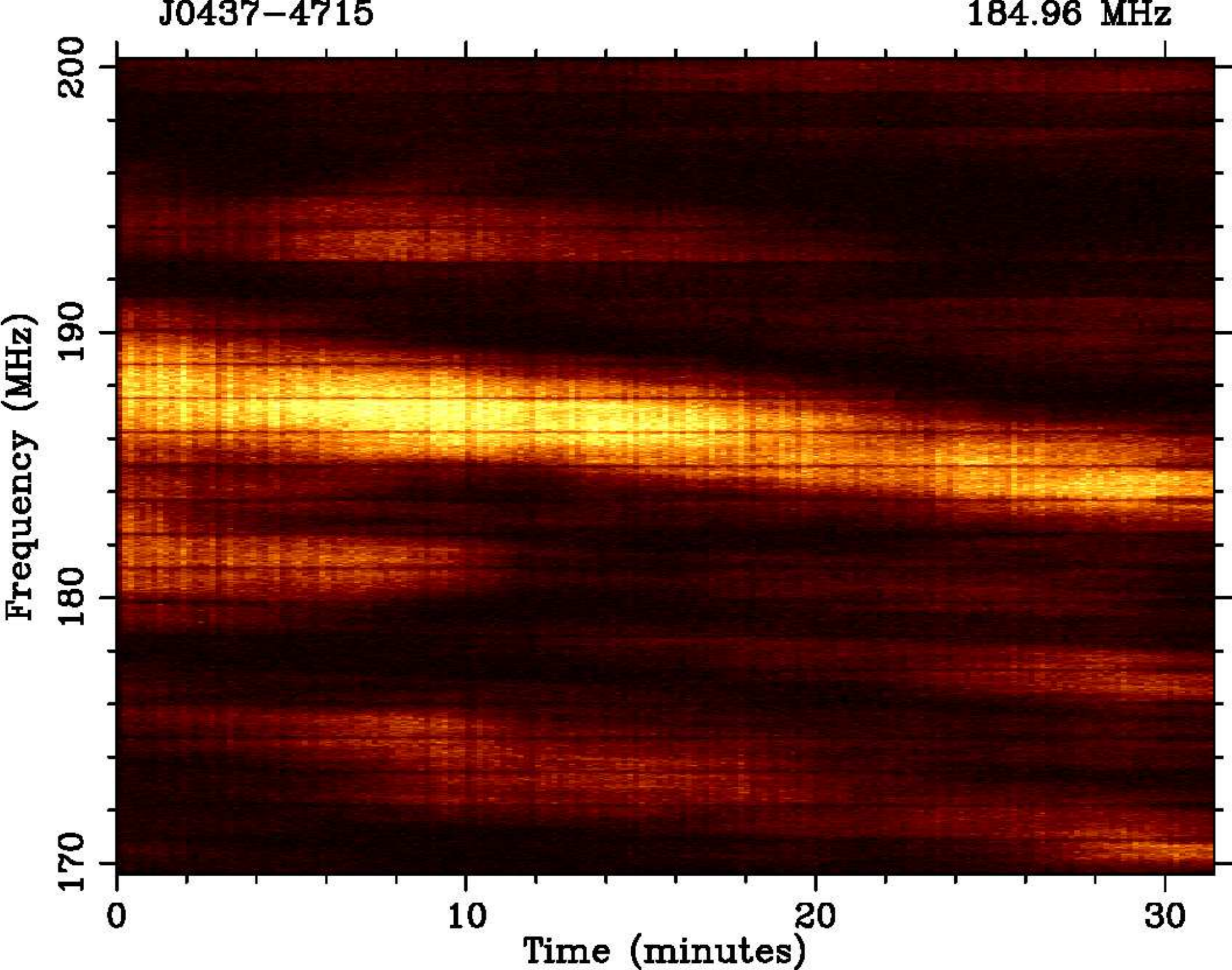}{0.5\textwidth}{(a)}
          \fig{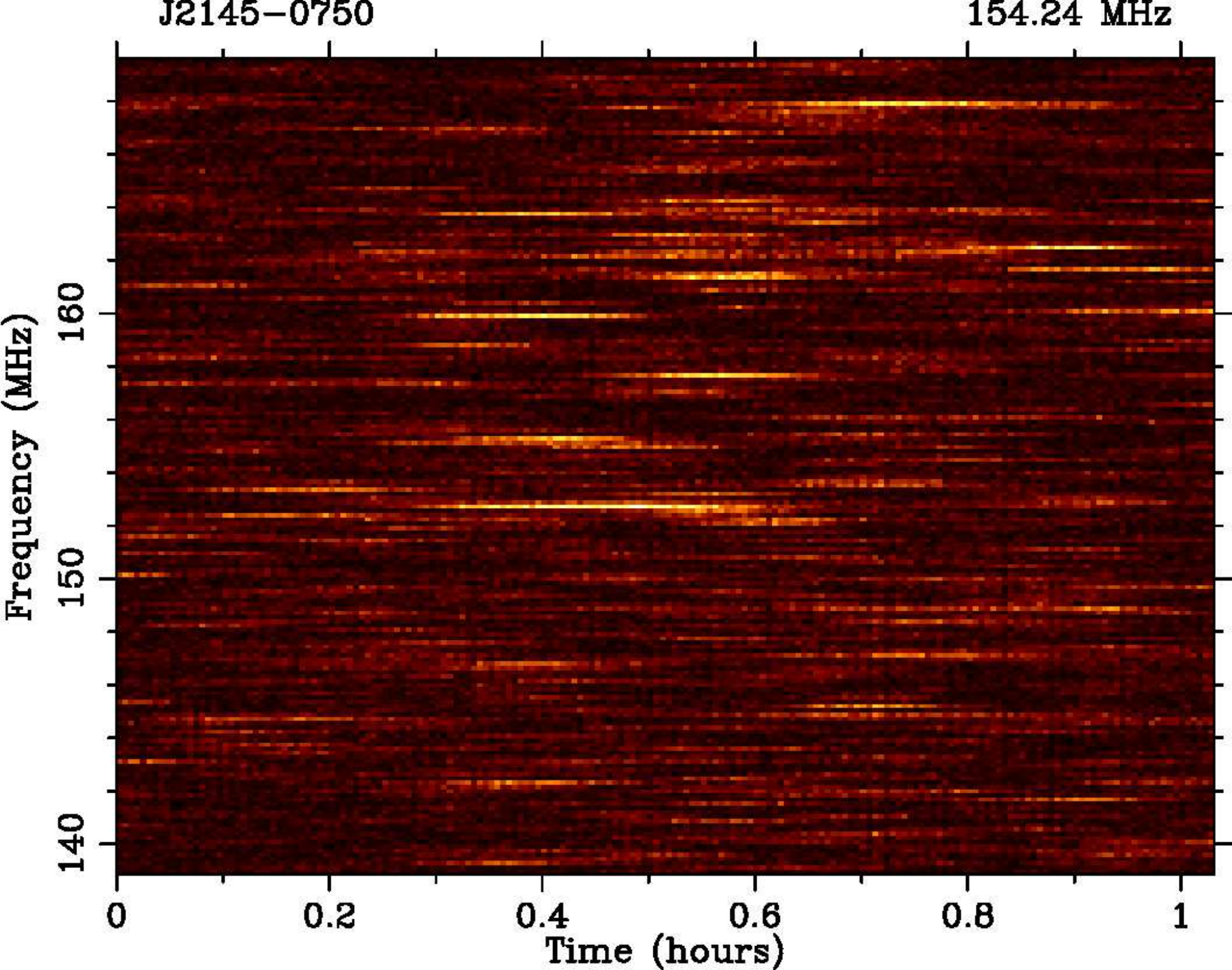}{0.5\textwidth}{(b)}
          }
\gridline{\fig{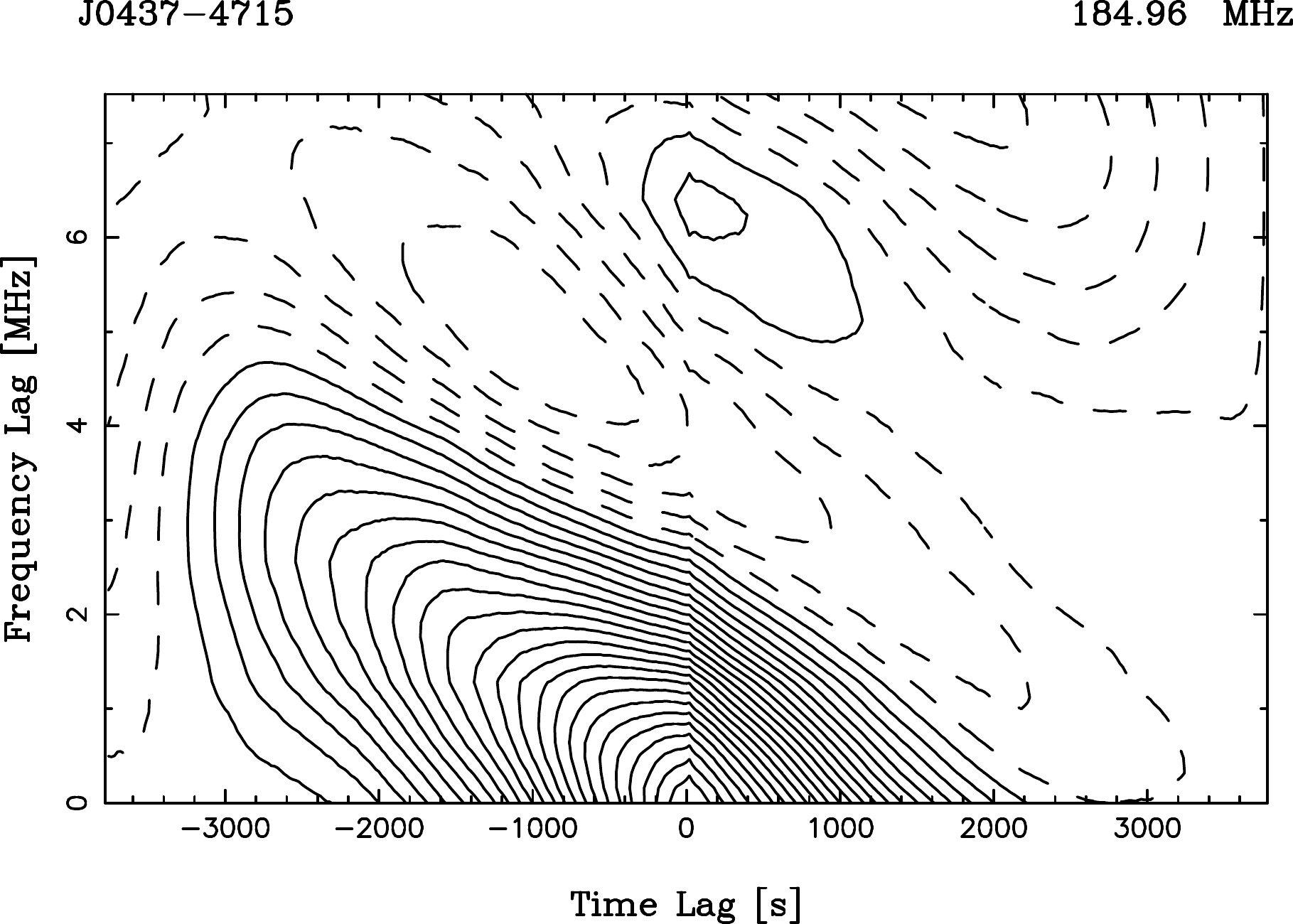}{0.475\textwidth}{(c)}
          \fig{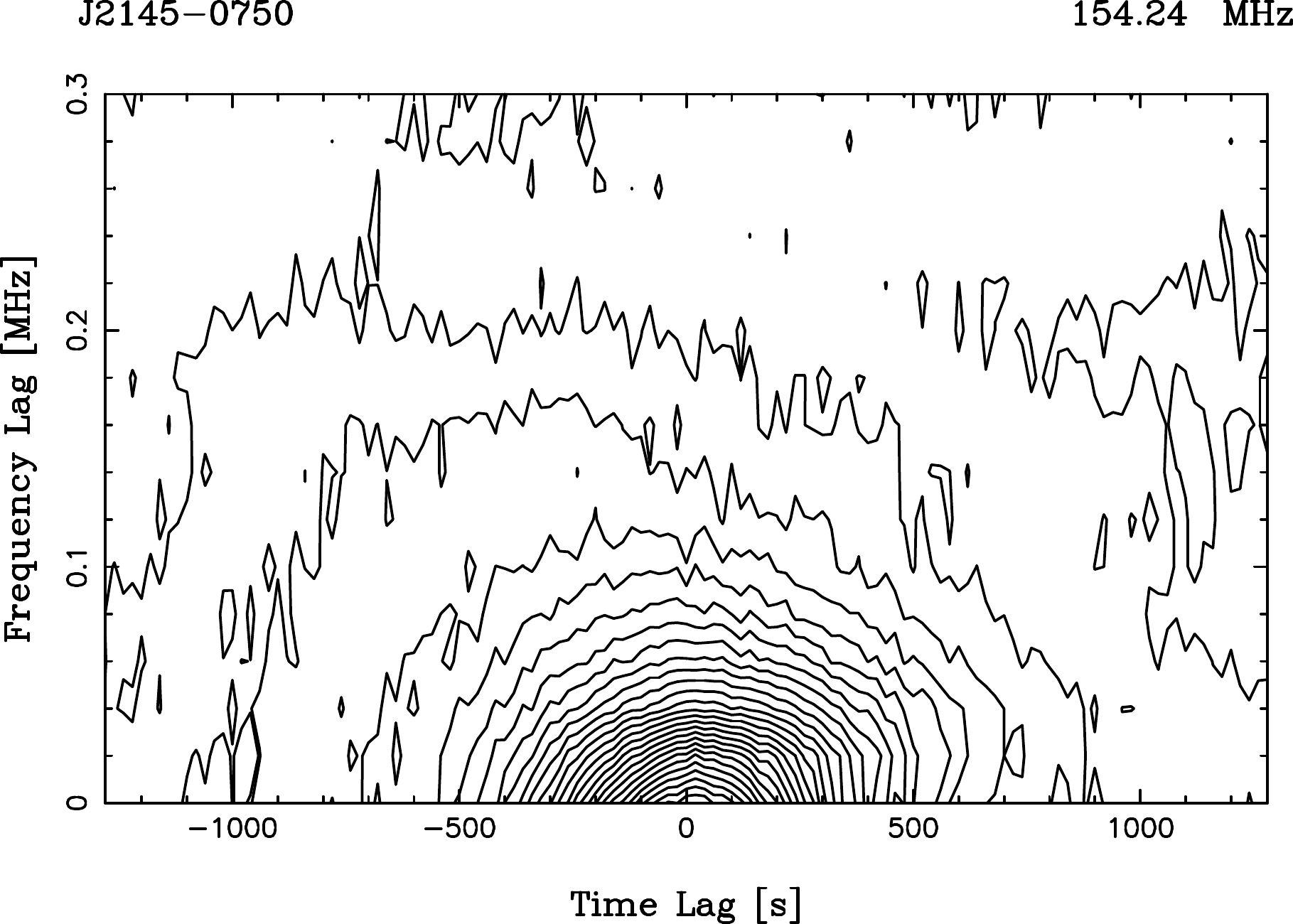}{0.475\textwidth}{(d)}
          }
\caption{Dynamic scintillation spectra of PSRs \psrnameone\ and \psrnametwo\ from MWA observations over a 30.72-MHz bandwidth (top panels), and the corresponding two-dimensional auto-correlation functions (bottom panels). Data resolutions are 10 s and 10 kHz in time and frequency, respectively. The horizontal stripes visible in the spectrum of \psrone\ are due to our 1.28-MHz coarse channelization in the MWA signal path; the effect is subdued in \psrtwo\ data, presumably due to much narrower sizes of its scintles (in frequency). 
\label{fig:f5}}
\end{figure*}

\begin{figure*}
\epsscale{1.1}
\plotone{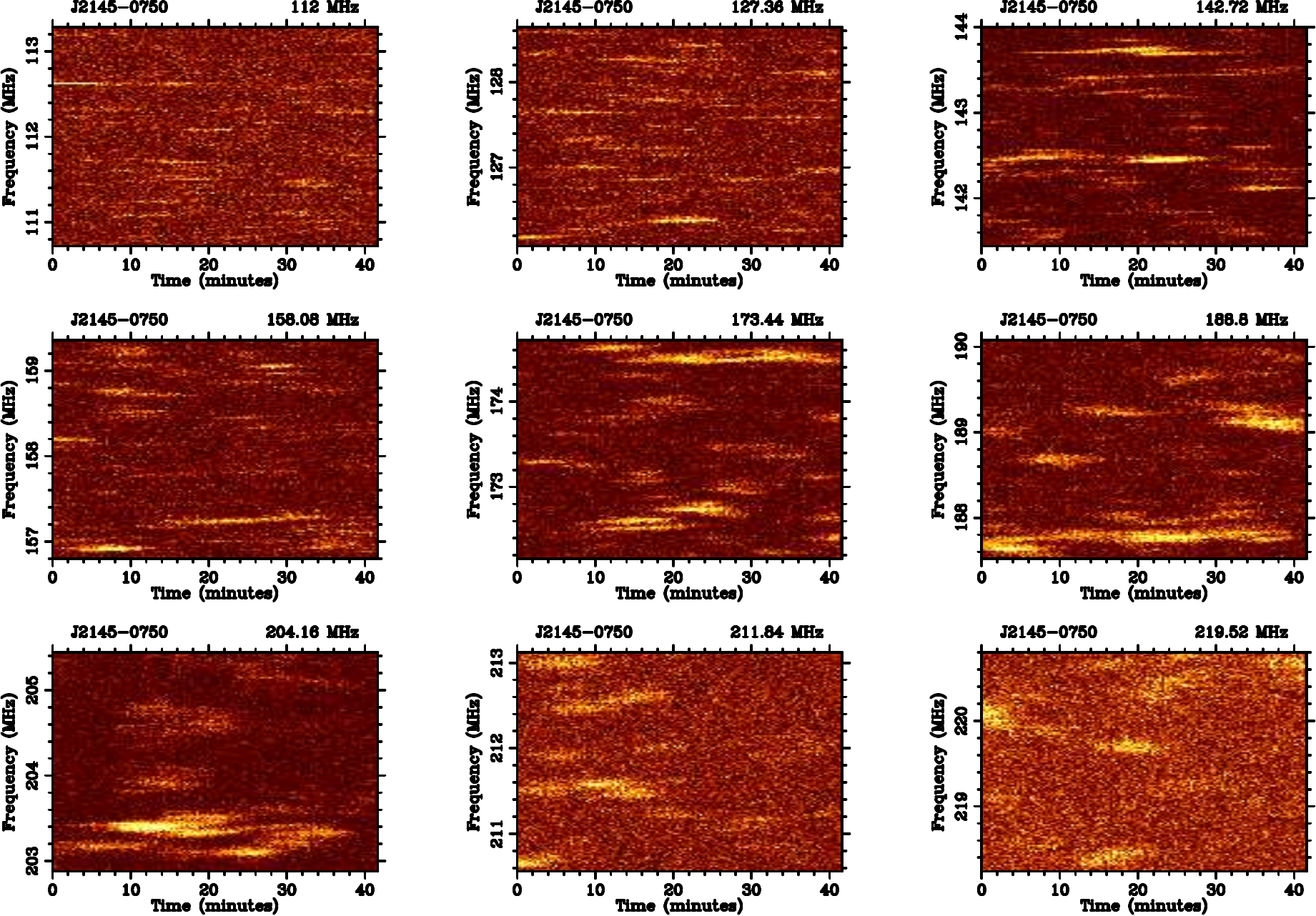}
\caption{
Dynamic scintillation spectra of \psrtwo\ at 9 different frequency bands across the MWA's 110-220 MHz range. Scintles are resolved throughout this range within the constraints of our observational parameters (i.e. 10-kHz spectral resolution and 2.56-MHz bandwidth). Data resolutions are 20 s and 20 kHz in time and frequency, respectively. 
\label{fig:f6}}
\end{figure*}

\section{Analysis and Results} \label{sec:res}

In this paper we limit the analysis to some basic properties that can be gleaned using the current capabilities of the MWA and EDA; specifically, pulsar detections in total intensity at frequencies spanning the 50--250 MHz range and scintillation studies that are possible with the 10-kHz spectral resolution of the VCS. More detailed analysis, e.g. full polarimetric pulse profiles and variability in DM and scintillation from an   ongoing monitoring project will be reported in a future paper. Even these exploratory observations provide useful insights into aspects such as the spectral evolution of the mean pulse profile and scintillation properties. We also measure pulsar flux densities to  investigate the spectral behavior at low frequencies; however, we caution that robust estimates will require data from multiple observations over long time spans, in order to average over variations caused by refractive modulation. 

\subsection{Pulse profiles} \label{sec:profs} 

Pulsar detections from our observations are summarized in Figs~\ref{fig:f1}-\ref{fig:f3}. In Figs~\ref{fig:f1}-\ref{fig:f2} we present MWA detections at multiple frequencies, along with those at higher frequencies (from Parkes), to illustrate the evolution in the mean pulse profile with frequency. 
MWA data are not coherently de-dispersed\footnote{The tied-array 
beam output is currently written out as full Stokes PSRFITS; the capability to generate higher time resolution voltage time series
is under development.}, and are therefore subject 
to residual dispersive smearing within the 10-kHz channel width. The resultant temporal smearing is relatively small (i.e. less than the 
100-$\mu$s native resolution of VCS) at frequencies down to 130 MHz for \psrone, and down to 195 MHz for \psrtwo, however becomes quite substantial at lower frequency bands; $\sim$0.5 ms at the lowest frequency bands of our detections, i.e., at 81 MHz for \psrone, and at 111 MHz for \psrtwo. 

For \psrtwo, dedispersion using the catalog DM of 8.998 \dmu results in a visible phase shift of approximately 0.11 phase turns across our 110--220 MHz range (see Fig.~\ref{fig:f2}). We therefore determined the DM from our own observations using the {\sc pdmp} utility of  the {\tt psrchive} package. Our measured DM = $9.004 \pm 0.003$ \dmu is significantly larger than the catalog value, and the implied DM excess ($\delta$DM=0.006 \dmu) is  in close agreement with that reported by \citet{dowell+2013} from their LWA observations at frequencies below 100 MHz. We then reprocessed our data at this refined DM. We however note that this excess is several times larger than the maximum DM variation of 0.001 \dmu that is reported from 6-yr span Parkes PTA data \citep{keith+2013}. Such a DM excess at low frequencies can possibly be attributed to chromatic DM, as theorised by \citet{cordes+2016}, but we defer further discussion on this topic until \S~\ref{sec:dm}. 

EDA data were coherently de-dispersed using the {\tt DSPSR} software package. Given the computational  constraints of the processing machine (32 GB of RAM), the data were processed using the convolving filter bank to simultaneously coherently de-disperse the signal while dividing the 327.68-MHz baseband stream into 32768$\times$10-kHz channels. This provides an effective time resolution of 100 $\mu$s, which is also the native time resolution of VCS data, and thus can facilitate meaningful comparison of pulse profiles. With a collecting area equivalent to 16 MWA tiles (i.e. 1/8th of the full array), the EDA is less sensitive (per unit bandwidth), however offers the advantage of obtaining virtually artifact-free pulse profiles. Our EDA detections of \psrone\ are shown in Fig.~\ref{fig:f3}, at frequencies in the 50--210 MHz range.\footnote{The dipole's response is known to be poor at frequencies \la 50 MHz, and the pulsar was comparatively weaker at frequencies \ga 200 MHz in this observation.}  

\begin{figure*}[t]
	\plottwo{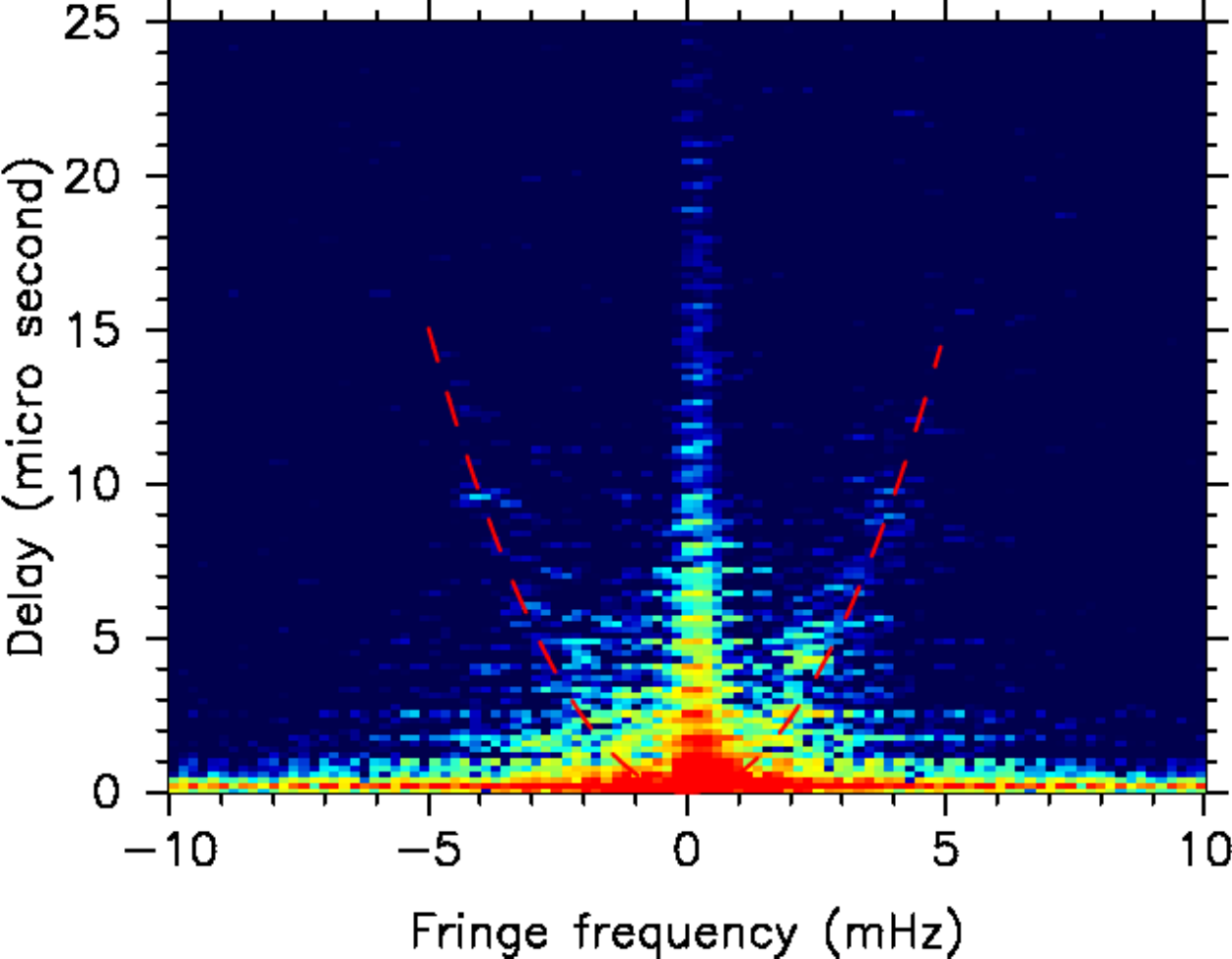}{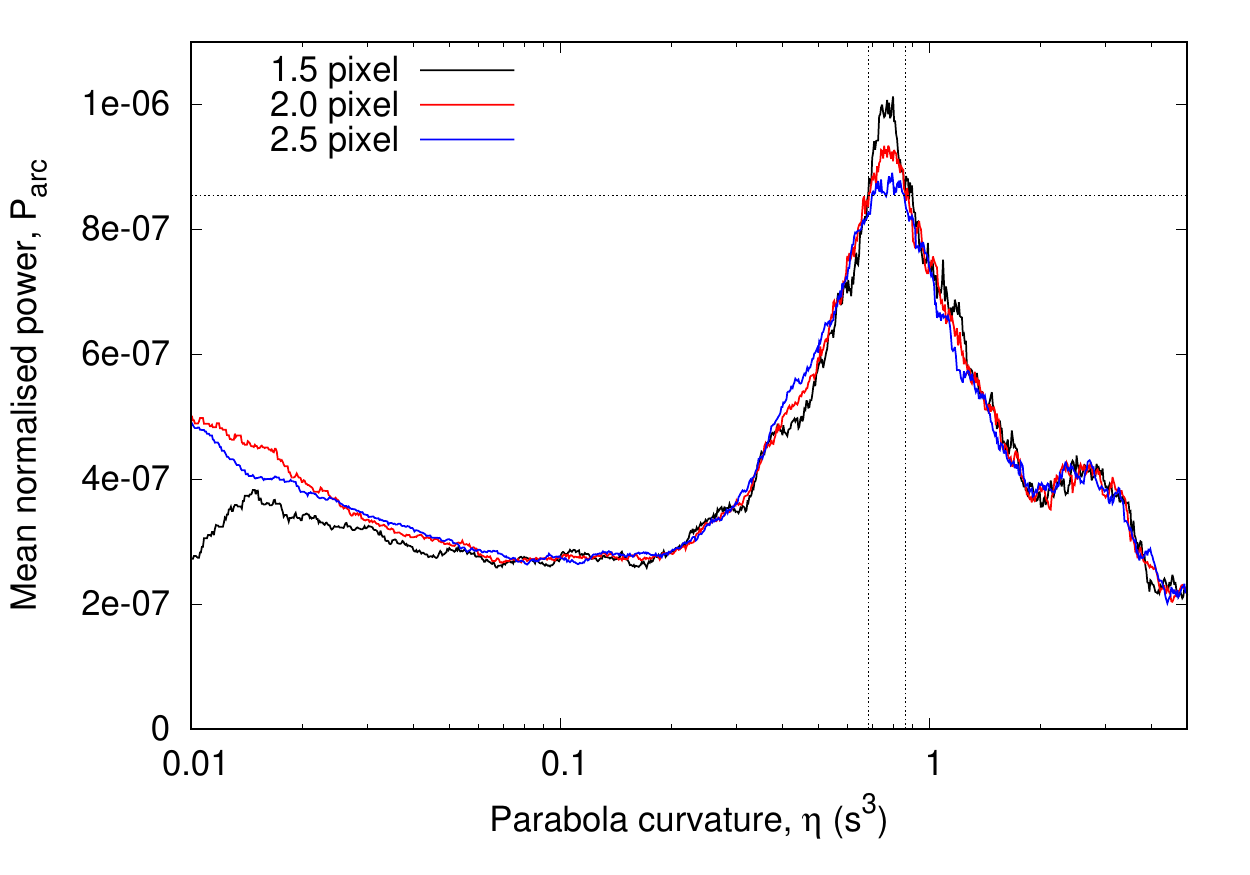}
	\caption{
		{\it Left}: Secondary spectrum of \psrone\ from the MWA observations shown in Fig.~\ref{fig:f5}(a); the resolutions  in the conjugate time (Doppler frequency) and conjugate frequency (delay) axes are 0.195\,mHz  and 0.195\,$\mu$s, respectively. The dashed red curve indicates the best-fit curvature estimated for the arc feature (see the text for details). 
		{\it Right}:  Mean arc strength $P_{\rm arc}$ against the curvature parameter $\eta$, computed for different thickness parameter ($p_{\rm d}$) of 
		the parabolic curve. The dashed horizontal line corresponds to 1$\sigma$ below the maximum value of $P_{\rm arc}$ (for the two-pixel curve), where $\sigma$ is computed based on the noise statistics estimated for a secondary spectral segment that is well outside the visible arc feature (see the text for details), whereas the dotted vertical lines correspond to the $\eta$ values where $P_{\rm arc}$ is 1$\sigma$ below the peak.
		\label{fig:f7}}
\end{figure*}

\subsubsection{Frequency evolution} \label{sec:freqevol} 

MSPs generally show complex pulse profiles, with a large number of components unlike most normal pulsars \citep[e.g.][]{dai+2015}. 
The degree of complexity often extends to polarization properties, including highly complex polarization position angle (PPA) variations across the pulse phase, which cannot be well described by the rotating vector model \citep[e.g.][]{kiriaki+1998}. Moreover, the emission is often seen to extend over a significant fraction of the pulsar period ($W_{10} \ga 0.5 \, P$, where $W_{10}$ is the width of the pulse measured at 10\% of the peak flux, and $P$ is the rotation period). Furthermore, the conventional radius-to-frequency mapping \citep{cordes1978,komesaroff+1970}, which often manifests as a considerable narrowing of the pulse with an increase in the observing frequency, is seldom seen  with MSPs \citep[e.g.][]{kramer+1998}. Overall,  MSP profiles have been quite challenging to model in terms of emission geometries,  
which may be due to multiple emission regions within the magnetosphere and pulse 
components originating at different locations \citep[cf.][]{gk1997,kiriaki+1998,kramer+1998,dyks+2010}. 

The recent work of \citet{dai+2015} has been very instructive in understanding the frequency evolution of MSP profiles. 
Using 6 years of PTA data from Parkes, they were able to perform phase-resolved spectral studies of MSP emission properties, including polarization variations across the pulse phase. \psrone\ is known to exhibit rather complex, and quite remarkable, evolution in its mean pulse shape \citep[e.g.][]{navarro+1997,bhat+2014}. Further, as noted by \citet{dai+2015}, the profile features in their high-quality Parkes data are generally consistent with our published MWA observations (at a frequency of $\sim$200 MHz), where a central bright component is flanked by multiple outer components (see Fig.~\ref{fig:f1}). Parkes data at very high signal-to-noise ratios (see Fig. A1 of \citealp{dai+2015}) also  reveal the emission extending out to  more than 85\% of the pulse period (i.e. $\sim$300$^{\circ}$ in longitude), which was first reported by \citet{yan+2011} (though also seen by \citealp{navarro+1997}). This is now also seen in our very high-quality MWA detection (S/N$\sim$3400) shown in Fig.~\ref{fig:f1} (the right panel), where the detectable emission extends to approximately 75\% of the pulse period (i.e. $\sim$260$^{\circ}$ in longitude). 

A comparison of MWA and Parkes data shows that the profile evolution of this pulsar is largely due to its highly complex spectral index ($\alpha$) variation across the pulse. As seen from the work of \citet{dai+2015}, $\alpha$ varies from $-0.8$ to $-2.2$ across the large emission window; while the leftward outer-edge component has a relatively flatter spectrum ($\alpha > -1.5$), the emission within a $\sim$0.2 phase range left of the central component shows a relatively steeper spectrum; $-1.7 > \alpha > -2.2$. The emission within a $\sim$0.2 phase range right of the central component  shows a more complex variation, whereby $\alpha$ varies from $-1$ to $-2.2$ rather smoothly, and marked by two prominent minima and maxima.
As seen from Fig. A1 in \citet{dai+2015}, the leading and trailing 
outer components
have generally steeper spectra, which may explain their rapid evolution at the MWA's frequencies. 

The pulsar profile is also known for its unusual double ``notch'' feature (see Fig.~\ref{fig:f1}; near phase $\approx$0.62 in the 438 MHz profile), besides a notch-like feature near the center; particularly in observations at frequencies in the 0.4-1.5 GHz range \citep{navarro+1997}. It appears that the disappearance of the central notch-like feature may be due to the complex spectral index changes. For instance, the rather abrupt change of $\alpha$ from $-2$ to $-1$ closer to the center (i.e. near  
phase 0 in \citealt{dai+2015}) may explain the ultimate low-frequency dominance of one component and disappearance of this central notch-like feature.

The disappearance of the double notch feature in MWA data is likely due to the limitation in our temporal resolution. The width of this feature is approximately 0.01 phase \citep{navarro+1997}, i.e. $\sim50\,\mu$s, and hence as such difficult to resolve with our 100-$\mu$s native time resolution.\footnote{The temporal broadening due to scattering is expected to be negligibly small (\la 1$\mu$s) at the MWA's 185 MHz.}
The location of this notch feature also happens to be near a local minimum in the spectral index variation, and the resulting non-uniform evolution of the components could also potentially lead to its disappearance in the MWA band. 
On the other hand, the leftward outer edge region of the profile (i.e. phase $<$0.3 in Fig.~\ref{fig:f1}) exhibits relatively flatter spectrum ($\alpha > -1.5$), which may also explain its reduced prominence at the MWA's frequencies. 

\psrtwo\ presents yet another interesting case. As evident from Fig.~\ref{fig:f2new}, the relative strengths of its leading and trailing components continue to decrease at lower observing frequencies and are reversed in the MWA band. Such a reversal was noted 
earlier by  \citet{kl96} in their observations at 102 MHz, and more recently in LOFAR data reported in \citet{vlad+2016}. This can be readily explained  given the results of \citet{dai+2015}, whose data we re-present in Fig.~\ref{fig:f2new}. 
As can be seen from their Fig. A23, the leading component of the pulse has $\alpha$ within the range $-1.5 > \alpha > -2$, whereas for the trailing component it is $-2 > \alpha > -2.5$.  This readily explains the observed reversal of the relative strength and its rapid evolution within the MWA band (Fig.~\ref{fig:f2}). The high time resolution of Parkes data also reveals finer structure, which is not resolved in MWA observations, largely due to the limited time resolution of VCS (100 $\mu$s) and non-negligible dispersive smearing (from our 10-kHz channels). Further, the precursor component that appears at 0.2 phase offset prior to the leading peak, is not seen in MWA data; this is quite perplexing considering its expected amplitude of $\sim$10-40\% of the leading peak, assuming the relatively steep spectral index ($\alpha \sim -2$) that was inferred by \citet{dai+2015}. This may suggest either a possible turnover of the precursor emission, or its plausible intermittent nature.  




\subsection{Scintillation} \label{sec:scnt}

The low frequency bands of the MWA are especially well-suited for scintillation studies of low-DM pulsars such as PSRs \psrnameone\ and \psrnametwo.  For instance, basic properties such as scintillation bandwidth $\nud$ and time scale $\tauiss$  can be deduced using well-established two-dimensional auto-correlation function (ACF) analysis  \citep[e.g.][]{gupta+1994,bhat+1999}, and are useful in estimating the integrated strength of scattering and scintillation velocities \citep[e.g.][]{cwb85,gupta1995}. Secondary spectral analysis can sometimes lead to the detection of parabolic scintillation arcs as we reported in \citet{bhat+2016} (see also \citealp{stine+2001}). Furthermore, within the constraints of our 10-kHz spectral resolution and the achievable sensitivity, it may also be possible to investigate the frequency scaling of scintillation parameters in some cases, given our multi-frequency sampling across large fractional bandwidths. 

\subsubsection{Dynamic spectra} 

The dynamic spectrum $S (f, t)$ is a two-dimensional record of pulse intensity in time and frequency. It is the most basic observable for scintillation analysis. Fig.~\ref{fig:f5}~and~\ref{fig:f6} show such spectra for PSRs \psrnameone\ and \psrnametwo\ obtained over 0.5 to 1 hr durations; over a continuous 30.72-MHz bandwidth (Fig.~\ref{fig:f5}), as well as 9 $\times$ 2.56 MHz bands for \psrtwo\ (Fig.~\ref{fig:f6}). Diffractive scintillation is seen as rapid, deep modulation of pulse intensity in time and frequency, whereas the drifting of intensity maxima (in the time-frequency plane) is generally attributed to refraction through the ISM. 
Drifting scintles can also be explained by the relative velocities of the Earth and pulsar projected onto the scattering screen; interference with the Doppler shifted scattered radio waves causes periodic temporal variability of pulse intensity.  These temporal variations give rise to drifting when combined with periodic spectral variability due to the temporal delay of the scattered rays \citep{wmsz04}.
Drifting scintles are most readily seen in observations of \psrone\ (Fig.\ref{fig:f5}), where the dynamic spectrum is dominated by a single bright scintle that extends over a time scale longer than our observing duration. As seen from these figures, scintles vary in their brightness, size (both in time and frequency) and orientation (i.e. drift rates), however average properties can be meaningfully characterized by computing a two-dimensional  ACF of $S (f, t)$ in  frequency and time lags: $\rho (\nu, \tau) = \langle S (f, t) \, S (f + \nu,  t + \tau) \rangle $, where $\nu$ and $\tau$ are the frequency and time lags, respectively. 


\begin{figure}[t]
	\plotone{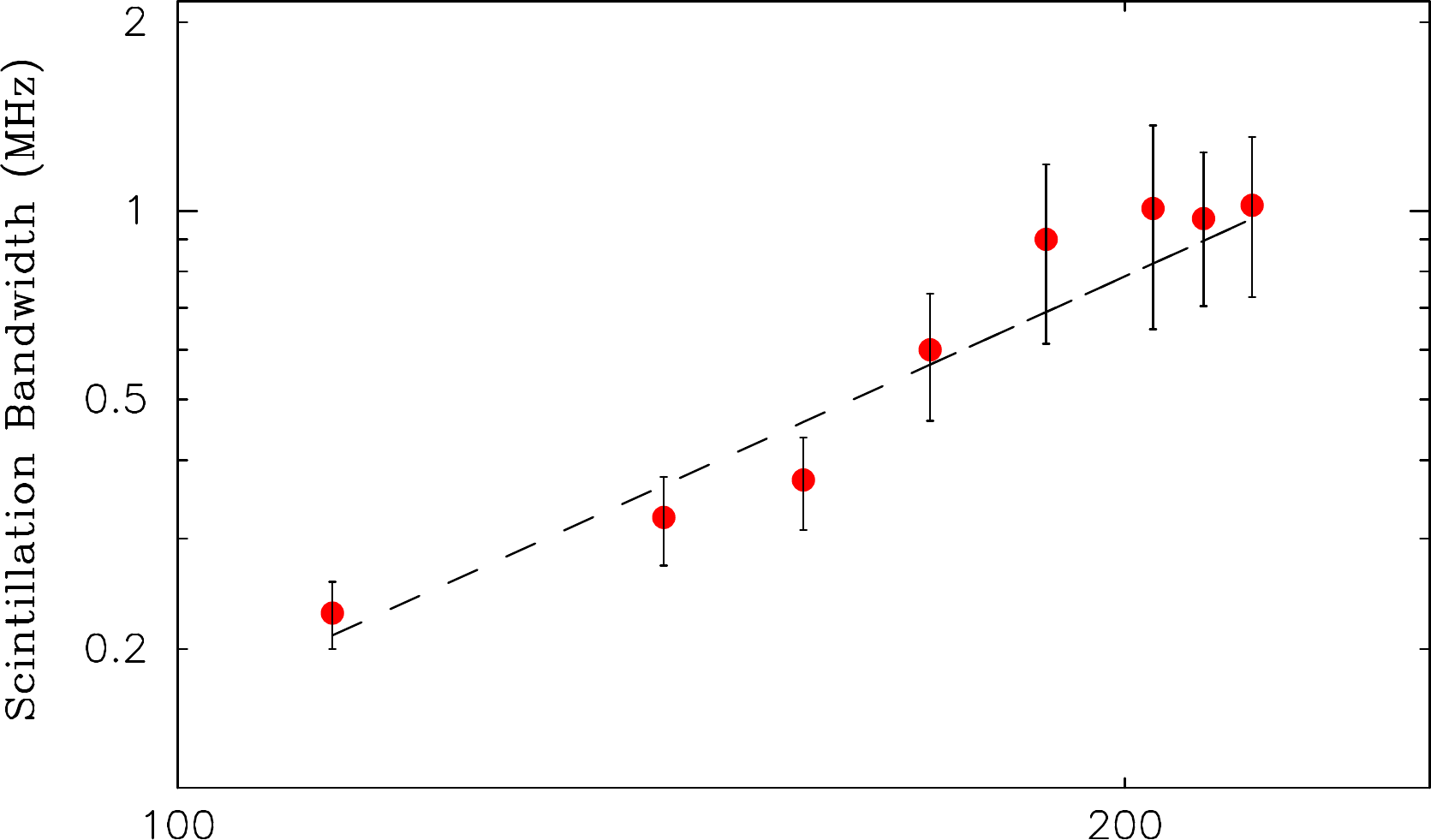}
	\plotone{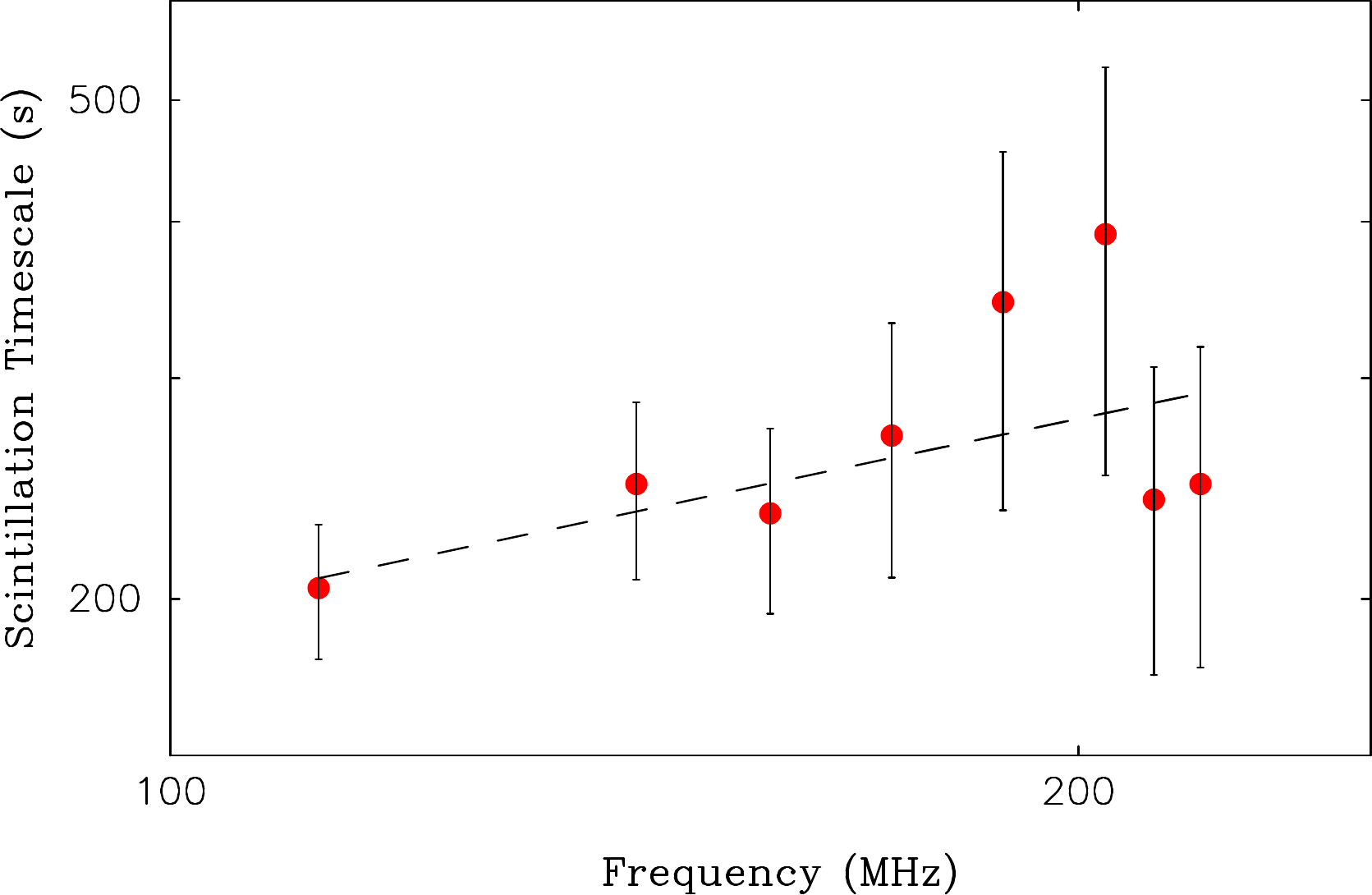} 
\caption{
Plots of scintillation bandwidth (\nud) and scintillation timescale (\tauiss) for \psrtwo\ at multiple frequencies within the MWA band. The uncertainties are due to the limited number of scintles, assuming a filling fraction $ f_{\rm d} $=0.5. The suggested scaling indices (i.e. the slopes of the best-fit lines) are $2.3\pm0.3$ for \nud, and $0.5\pm0.3$ for \tauiss. 
\label{fig:f9}}
\end{figure}


\begin{deluxetable*}{ccCcccc}[t]
\tablecaption{Observational parameters with the EDA  \label{tab:eda}}
\tablecolumns{6}
\tablenum{2}
\tablewidth{0pt}
\tablehead{
\colhead{PSR} &
\colhead{MJD range} &
\colhead{$ N _{\rm obs}$} &    
\colhead{$ T _{\rm span}$} &  
\colhead{$ T _{\rm obs}$} & {Data length} & 
\colhead{Frequency range \tablenotemark{a} } \\
\colhead{} & \colhead{} & \colhead{} & 
\colhead{(days)} & \colhead{(seconds)} &
\colhead{(seconds)} & \colhead{(MHz)}
}
\startdata
\psrnameone & 57743-57879 & 6 & 136 & 300 & 1200 & 50-220 \\   
\psrnametwo &57851-57951 & 9 & 100 & 288 & 2500 & 70-220 \\
\enddata
\tablenotetext{a}{The frequency range of pulsar detection. The lowest frequency for processing depends on the pulsar DM (see the text).}
\tablecomments{$N _{\rm obs}$: the number of observations; 
$ T _{\rm span} $: the time span of observations;  
$T _{\rm obs}$: the observing duration per session.}
\end{deluxetable*} 

The ACFs computed in this manner are shown in the lower panels of Fig.~\ref{fig:f5}. The characteristic  widths in frequency and time, i.e. scintillation bandwidth \nud and timescale \tauiss, respectively, can be determined by fitting a two-dimensional elliptical Gaussian function to these ACFs.  Following  the published literature  \citep{gupta+1994,bhat+1999,wang+2005}, we fit a functional form 
$\rho_g\,(\nu,~\tau) = C_0 \, {\rm exp}  [ - ( C_1 \nu^2 + C_2 \nu \tau + C_3 \tau^2 ) ] $, to yield characteristic 
scales: the decorrelation bandwidth, 
$\nu_{\rm d}=({\rm ln\,2}/C_1)^{0.5}$, measured as the half-width at half-maximum of the correlation peak; 
scintillation timescale, $\tauiss=(1/C_3)^{0.5}$, measured as the $1/e$ width of the correlation maximum; and the drift rate, 
$dt/d\nu=-(C_2/2\,C_3)$. 
In addition, a ``drift-corrected" scintillation bandwidth can also be estimated, i.e.   
$ \nudc = ({\rm ln\,2})^{0.5} \, ( C_1 - { C_2 ^2 / 4 \, C_3 } )^{-0.5} $, to reduce the  underestimation bias resulting from significant refraction typically seen in our data. 
The parameters determined in this manner are tabulated in Table~\ref{tab:scnt}, along with some physical properties that can be derived, namely: the mean turbulence strength \Cnsq, the strength of scattering $u$, and the scintillation pattern speed \Viss. Further, the diffractive and refractive scattering angles (\thetaref and \thetadiff) can be estimated from the measurements of decorrelation bandwidth and the drift rate \dtnu, respectively,  under the assumption of thin screen scattering model, which is generally employed in the context of scintillation \citep[cf.][]{cordes+1986,gupta+1994,bhat+1999}. 

For \psrone, we measure $\nud = 1.4 \pm 0.4 $ MHz\footnote{The large uncertainty is due to small number statistics arising from our limited number of scintles (cf. \citealp{bhat+2014}).}, which is consistent with that reported in an earlier publication \citep{bhat+2014}. 
As we summarised in that paper, there are striking discrepancies between previously published measurements, which may be attributed 
to the choice of sub-optimal observing parameters in much of the earlier work. The mean turbulence strength $\Cnsq = 8 \times 10^{-5}$
\cnsqunits, is the second lowest (after PSR B0950+08; see \citealp{pc92}) measured amongst all known pulsars, making \psrone\ one of 
the most weakly scattered pulsars. Any timing perturbations resulting from scattering can therefore be safely ignored at its timing frequencies ($\sim$1-2 GHz). 

The measured drift rate $\dtnu$\,is almost twice as large as our previous observations and of opposite sign, though this can be attributed to time-varying refraction that is commonly seen toward many of the low-DM pulsars \citep{wang+2005,bhat+1999}.
The refractive scattering angle \thetaref $\sim$ 0.2 mas is of comparable magnitude still, and it is the combination of a large \nud, the pulsar's high proper motion (\Vmu = 100 \velu), and a long time scale (\tauiss = 613$\pm$165 s) that gives rise to visibly prominent drift patterns that we see in the MWA band. Such prominence of refraction would generally imply the existence of discrete density structures along the line of sight, which can also give rise to parabolic scintillation arcs that we discuss later in \S~\ref{sec:arcs}. 
Variability of the measured drift rate can also be explained by the orbital motions of the Earth and pulsar; as their relative velocities projected onto the thin screen vary with time, so does the Doppler shift of radio waves scattered off asymmetric structures in the scattering screen.

For the \psrtwo\ observations presented in Fig.~\ref{fig:f5} (over a 30.72-MHz bandwidth centred at 154.24 MHz), we measure $ \nud = 27 \pm 4$  kHz and \tauiss = $300 \pm 45$ s. Thus the scintles are barely resolved (in frequency) with our 10-kHz spectral resolution with the VCS. The derived value for \Cnsq\ is nearly twice that of  \psrone, making the sight line of \psrtwo\ comparatively less anomalous.

The scintillation velocity \Viss can be derived from the measurements of \nud and \tauiss; e.g. in the simplest case where the scattering medium is approximated as a thin screen located at a certain distance between the pulsar and the observer, \Viss is given by \citep[cf.][]{gupta+1994}
\be
\Viss ~ = ~ A_{\rm ISS}  \, { ( { \nu _{\rm d} \, D \, x } )^{1/2}   ( { \nu _{\rm obs}\,  \tau _{\rm iss}} )^{-1} }, 
\label{eq:viss} 
\ee
where $D$=$\Dos$+$\Dps$ is the pulsar distance, and $x=\Dos/\Dps$, i.e. the ratio of the distances from the screen to the observer ($\Dos$) and from the screen to the pulsar ($\Dps$). The constant, $A_{\rm ISS}$, relates the time scale \tauiss to the velocity, for which \citet{cr1998} derive a value of $2.53\times10^4$ \velu\ for a Kolmogorov turbulence spectrum and homogeneously distributed medium (with $\nu _{\rm d}$ in MHz, $\nu _{\rm obs}$ in GHz, $\tau _{\rm iss}$ in s, and $D$ in kpc), and \citet{gupta+1994} derive $ A_{\rm ISS} = 3.85 \times 10^4 $ \velu\ for a single asymmetrically located thin screen. 

The estimated scintillation velocity  \Viss = $99 \pm 18$ \velu (including the uncertainty in the pulsar's distance, $D$=$530\pm60$ pc) is three times larger than the measured proper motion of  33 \velu for this pulsar \citep{reardon+2016}, which may suggest an asymmetric location for the underlying scattering screen. 

\subsubsection{Scintillation arcs in \psrone\ revisited} \label{sec:arcs}

The two-dimensional power spectrum  of the dynamic spectrum $S (f,\,t)$ is referred to as the secondary spectrum,  $S_2(\fnu,\,\ft)\,=\,| S ^{\dagger}(\nu,\, t)|^2$ (where $\dagger$ indicates two-dimensional Fourier transform), and is a powerful technique that captures interference patterns produced by different points in the image plane \citep[e.g.][]{stine+2001,wmsz04,cordes+2006}.  The  ``fringe rates"  (in time and frequency),  \ft\ and \fnu,  are essentially the fringe frequency and delay parameters,  where \fnu is a measure of the differential time delay between pairs of rays, and \ft is the temporal fringe frequency. Interference between scattered wavefronts from the origin and pairs of points along an axis in the direction of the net velocity vector  \Veff\ produces parabolic scintillation arcs, which can be represented by $\fnu=\eta\ftsq$. The parabolic arcs are thus essentially a natural consequence of small-angle forward scattering. The fringe frequency and delay parameters \fnu\ and \ft\ can be related to the curvature of the arc ($\eta$); following \citet{cordes+2006}, 
\be
\eta =  { D\,s\,(1-s)\,\lambda^2 \over 2\,c\,\Veffsq \, {\rm cos}^2 {\theta} }  
\label{eq:eta}
\ee
where $D$ is the pulsar distance, $s$ the fractional distance of the screen from the source (i.e. $s=\Dps/D$, and $s=0$ at the source), 
$D_{\rm eff} \equiv D\,s\,(1-s) $ the effective distance to the screen, and $\theta$ the angle between the net velocity vector \vecVeff\ and the orientation of the scattered image. The quantity \Veff\ is the velocity of the point in the screen intersected by a straight line from the pulsar to the observer, and is the weighted sum of the pulsar's binary and proper motions (\Vbin\ and \Vmu\, respectively), and the motion of the screen and the observer (\Vscreen\ and \Vobs\ respectively). Its transverse component 
\vecVeffperp (cf. eq. 3 in \citealp{bhat+2016}) determines the measured timescale \tauiss.  
Thus, the measurement of $\eta$ can be used to determine the location of the scatterer, when all the contributing terms are precisely known.\footnote{In general,  $s$ can be determined to within a pair of solutions, however the degeneracy can be often resolved by making use of other measurements or constraints (e.g. estimates of \Viss and \Vmu).}. 

In a previous publication \citep{bhat+2016} that reported our first observations of scintillation arcs in \psrone, we used the measurement of $\eta$ and the knowledge of the pulsar's precise orbital and astrometric parameters, to determine the location of the underlying scatterer to be at $115\pm3$ pc. Interestingly, this was consistent with the location inferred from independent observations from Parkes made two weeks later at a frequency of 732 MHz.

Fig.~\ref{fig:f7} shows the secondary spectrum from our new observations of \psrone\ (for the dynamic spectrum shown in Fig.~\ref{fig:f5}); the resolutions in fringe frequency and delay axes are 0.195 mHz and 0.195 $\mu$s, respectively (i.e. identical to those in \citealt{bhat+2016}). The parabolic arc features are consistently seen across different frequency segments of these new observations, albeit somewhat less prominently in the upper one-sixth of the band, where the pulsar is visibly dimmer compared to the remainder of the observing band. Furthermore, the arc feature is also visibly stronger than our previous detection, albeit more diffuse in appearance, which is reminiscent of a filled parabola \citep[e.g.][]{ps2006}. Following a feature-extraction technique as described in \citet{bhat+2016}, we compute an ``arc strength" parameter $\Parc$, given by 
\be
P _{\rm arc} ( \eta )  = { 1 \over N } \sum _{i=1} ^N S _{2} ( \eta \ftisq, \, \fti ). 
\ee
This method, which 
can be compared to
the one-dimensional generalized Hough transform, is expected to be more robust, particularly when the arc feature is limited by signal-to-noise, which is the case with our detections. The summing procedure  is performed along the points of arc outside an excluded low-frequency noise region, and out to delays beyond which little power is detectable (i.e. $i$=1...$N$, where $N$ corresponds to $\fnu$=20 $\mu$s). Further, in order to better account for the visibly larger diffused arc feature, we consider a slightly modified parameterisation for the ``thickness" of the parabola, $\pdist$, which is essentially the distance (in pixel units) from the arc curvature that is used for the computation of $S _{2}$; e.g. $\pdist=2$ means that only those spectral points lying within 2 pixels of the parabola (either horizontal or vertical) are included in the sum. This is expected to yield a more robust parameterisation, particularly when the arc feature is more diffuse in nature. 

The right panel of Fig.~\ref{fig:f7} shows our estimated $P _{\rm arc} ( \eta )$ for the secondary spectrum shown in the left panel of this figure. The curve appears to be more broadly-peaked than that of our earlier detection, despite its visibly higher degree of prominence. 
A closer examination of the secondary spectrum also reveals the arc feature is more prominent in the right half. The best-fit value of the curvature of the arc $\eta$, which corresponds to the maximum value of \Parc, is $\eta = 0.79 \pm 0.03$, for the case where 
$\pdist=1.5$, however the estimate is somewhat more uncertain if we consider the more broadly-peaked curve for $\pdist=2.5$ 
(i.e. a thicker parabola). Nevertheless, all three curves peak at similar values of $\eta$.\footnote{The computation of \Parc is restricted 
to $\ft > 0$ where the arc feature is prominent.} 
The implied distance to the
screen (after accounting for the expected \Veff at our observing epoch; see \citet{bhat+2016} for details),  is $s=0.26\pm0.02$ (or $115\pm 3$ pc), which 
is in excellent agreement with our previously reported measurements. This stability of the $\eta$ value re-affirms the presence of a compact scatterer in the line of sight to the pulsar. 

As seen from eq. 2, the inferred location $s$  depends also on the orientation $\theta$ of the scattered image relative to \vecVeff (and is generally an unknown).   However, since the 3-dimensional sky geometry of the pulsar is precisely known \citep{willem+2001}, it is indeed possible to calculate the projection of the vector velocity \vecVeff relative to the scattering screen (assuming it is on the plane of the sky). The effective velocity is still dominated by the proper motion of the binary system; and for the two epochs of MWA observations, incidentally, the projection angles differ by only 2.6$^{\circ}$. Therefore, any resultant changes in $\eta$ are not easily measurable given our uncertainties. 

\begin{figure*}
\epsscale{1.123}
\gridline{\fig{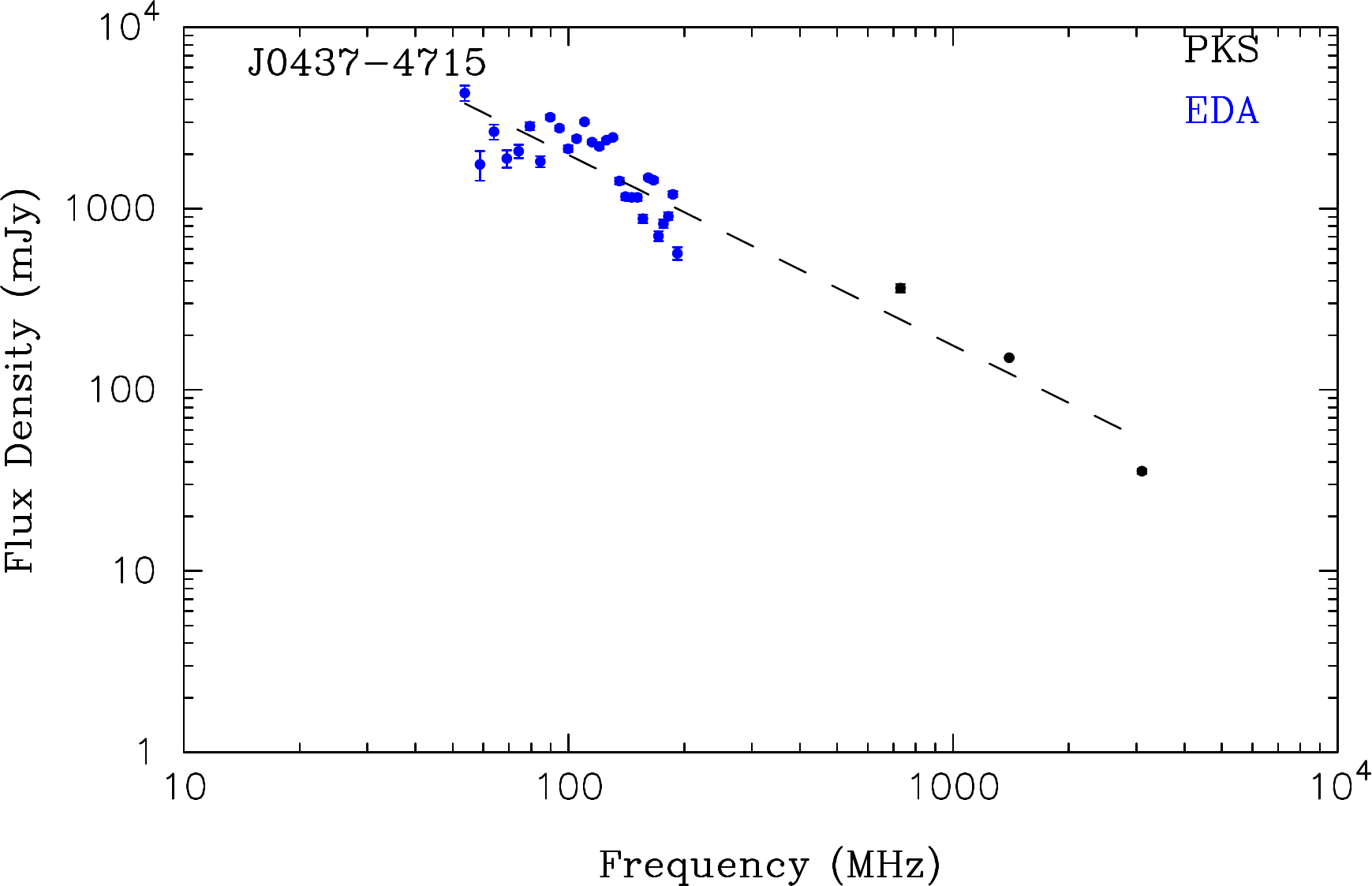}{0.5\textwidth}{(a)}
          \fig{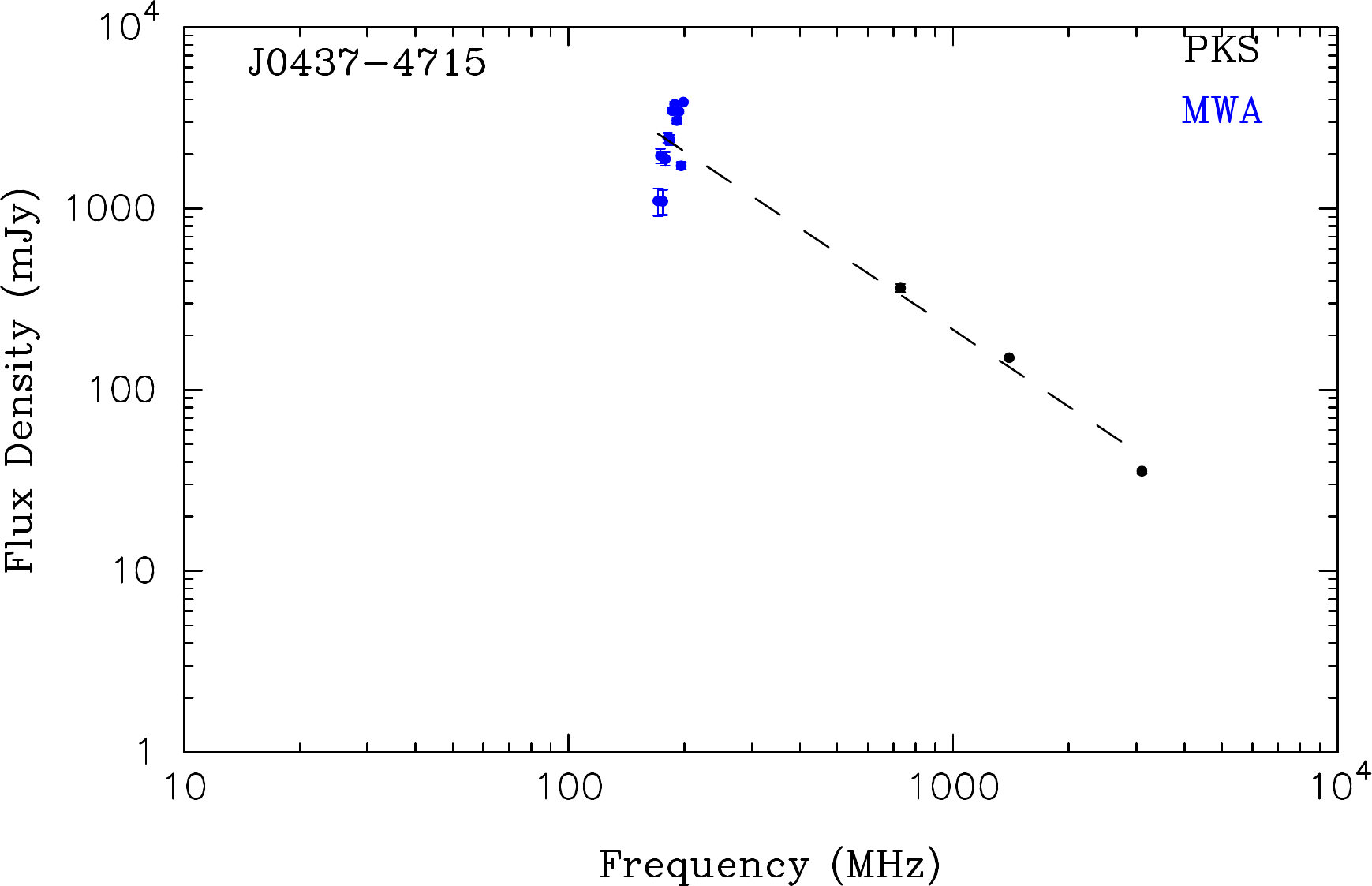}{0.5\textwidth}{(b)}
          }
\gridline{\fig{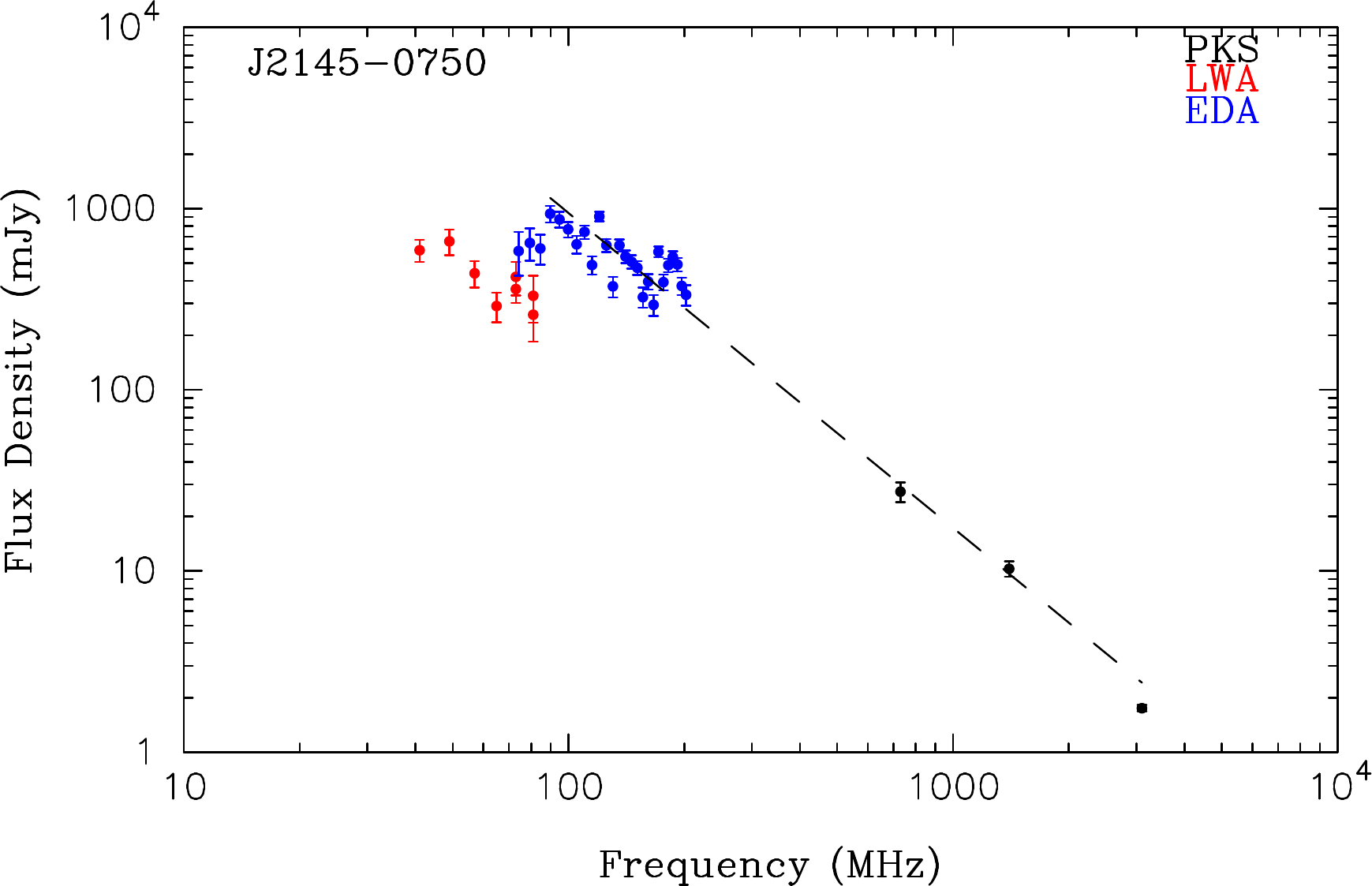}{0.475\textwidth}{(c)}
          \fig{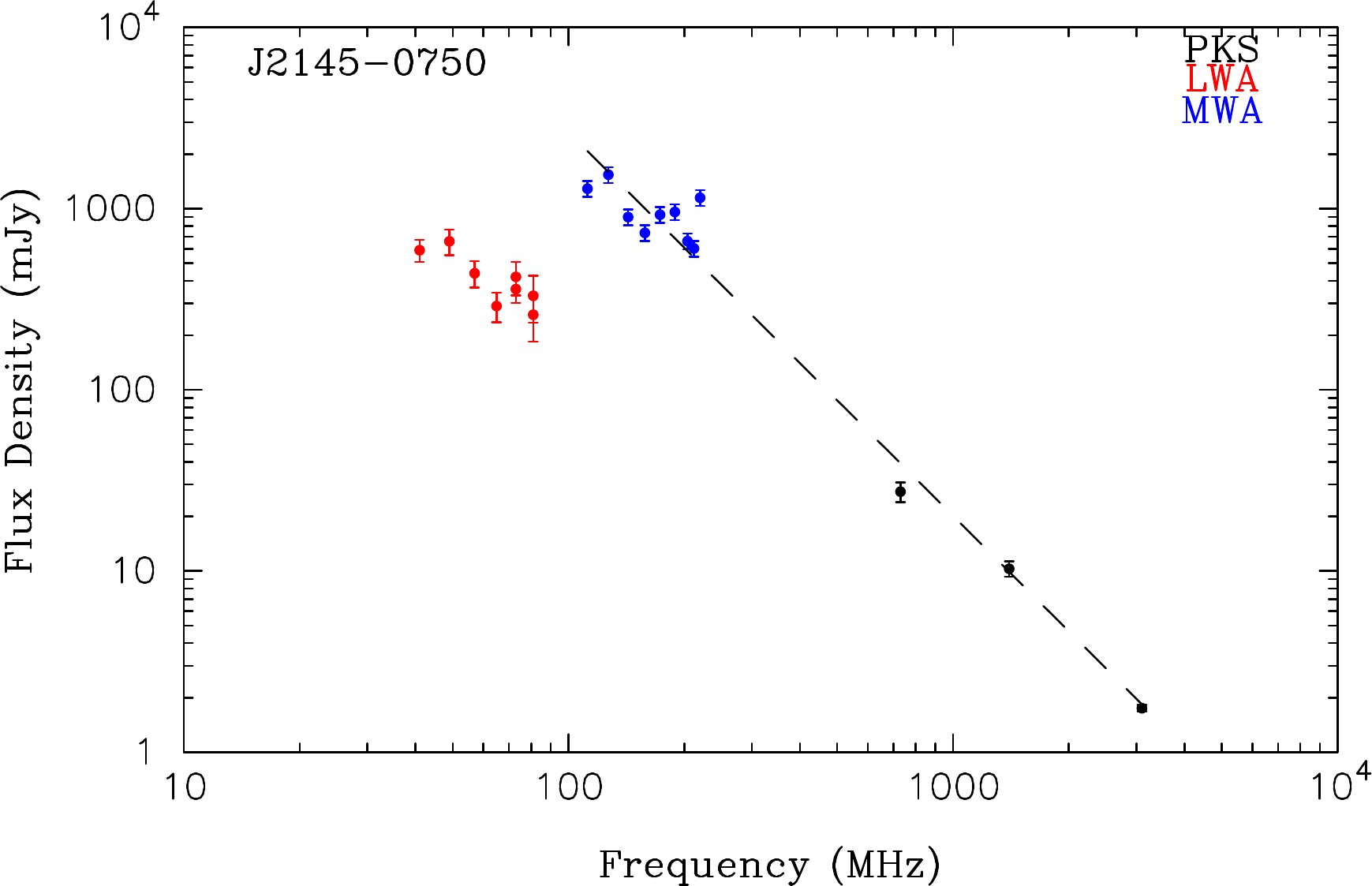}{0.475\textwidth}{(d)}
          }
\caption{
Flux density measurements of PSRs \psrnameone\ and \psrnametwo\ at frequencies spanning from 30 MHz to 3 GHz. {\it Upper panels}: EDA and MWA measurements of \psrnameone\ along with Parkes (PKS) measurements; {\it lower panels}: EDA and MWA measurements of \psrnametwo\ along with LWA and Parkes measurements. LWA data are from Dowell et al. (2013), whereas Parkes measurements are from Dai et al. (2015), and are more reliable as they are obtained from 6-yr data span. EDA measurements are averaged over multiple observations spanning several months (see Table 2). The best-fit spectral index estimates are summarized in Table 4. 
\label{fig:f8}}
\end{figure*}

\subsubsection{Frequency scaling of scintillation parameters}  \label{sec:freq} 

Simultaneous multi-frequency observations such as those presented in this paper can also be used, in principle, to investigate the frequency scaling of scintillation parameters. Theoretical treatments based on a Kolomogorov-type turbulence spectrum for electron density fluctuations in the ISM predict $\nud \propto \nu^{4.4}$ and $\tauiss \propto \nu^{1.2}$ \citep[e.g.][]{cordes+1986}. However, departures from these theoretical scalings have been seen toward a number of objects and, on a global scale, observations indicate a shallower scaling (e.g. $\nud \propto \nu^{3.9\pm0.2}$; \citealt{bhat+2004}).  
An important caveat is, at low frequencies, long-term (refractive) modulations in \nud and \tauiss can bias the measured value at a given epoch of observation \citep[e.g.][]{gupta+1994,bhat+1999}, and ideally observations spanning many refractive cycles are needed for a reliable estimation of the frequency scaling index. 

Notwithstanding this, we attempted the related analysis using our multi-band observations. \psrone\ data were not suitable for this purpose, as the expected $\nud\ga$ 1.28 MHz bandwidth for a large subset of our data (at frequencies \ga 150 MHz). However, our measurement of $\nud \sim$ 35 kHz for \psrtwo\ in the 140-170 MHz band (see Fig.~\ref{fig:f5}) suggests that  \nud is measurable within the constraints of our 10-kHz spectral resolution (and 2.56-MHz sub-bands) at frequencies \ga 140 MHz. 

Fig.~\ref{fig:f6} shows the dynamic spectra from our \psrtwo\ observations shown in  Fig.~\ref{fig:f2}.  As seen from this figure, scintles are resolved within our observing parameters, i.e. 10 kHz \la \nud \la 2.56 MHz over the 112-220 MHz frequency range of our observation. This demonstrates the value of the MWA's observable frequency range for studying scintillation parameters as a function of frequency. However, because of the limited S/N, a full 2-D correlation analysis proved less meaningful, and we therefore adopted a relatively simpler analysis, whereby we make use of 1-dimensional cuts of the 2D ACF along $\nu=0$ and $\tau=0$ axes to estimate $\nud$ and \tauiss (by fitting for a Guassian function). 

Our estimates obtained in this manner are shown in Fig.~\ref{fig:f9}, and the derived scaling indices are $2.3\pm0.3$ for \nud and $0.5\pm0.3$ for \tauiss, i.e., much shallower than we expect based on the empirically determined scaling or the theoretical predictions. However this is perhaps less surprising as our estimates are derived from a single epoch of observations and the refractive modulation may not necessarily be correlated across our large fractional bandwidth. This is probably best alleviated by multiple similar observations that span a large number of refractive cycles, and by employing more optimal observing strategies (e.g. fewer sub-bands with larger sub-bandwidth to allow 2D correlation analysis and more reliable estimates). 



\begin{deluxetable*}{ccCccccccc}[t]
\tablecaption{Scintillation parameters from MWA observations \label{tab:scnt}}
\tablecolumns{10}
\tablenum{3}
\tablewidth{0pt}
\tablehead{
\colhead{PSR} &
\colhead{\nud} &
\colhead{\tauiss} &    
\colhead{\dtnu} &   
\colhead{\Viss} &  
\colhead{\Vmu} & 
\colhead{\Cnsq} & 
\colhead{\thetadiff} &
\colhead{$ | \thetaref | $} &
\colhead{\sd \tablenotemark{a}}\ \\
\colhead{} &
\colhead{(MHz)} &
\colhead{(s)} &    
\colhead{(${\rm s\,MHz^{-1}}$)} &   
\colhead{(\velu)} &  
\colhead{(\velu)} & 
\colhead{(\cnsqunits)} & 
\colhead{(mas)} &
\colhead{(mas)} &
\colhead{}
}
\startdata
\psrnameone & $1.43\pm 0.4$ 	& $613\pm 165$	& $-181$ 	& $114\pm 36$	& 100	& $8 \times 10^{-5}$ 	& 0.8	& $0.2$ & 0.58 \\
\psrnametwo & $0.027\pm 0.004$	& $300\pm 45$	& 1080	& $99\pm 18$		& 33		& $1.4 \times 10^{-4}$ 	& 3		& 0.2 & 0.09 \\
\enddata
\tablenotetext{a}{The sparseness parameter that quantifies the time-frequency occupancy of scintles (see the text for details).}
\tablecomments{table comments}
\end{deluxetable*}

\subsubsection{Sparseness in dynamic spectra} \label{sec:sparse}


The dynamic spectra of \psrtwo\  in Fig.~\ref{fig:f6} are marked by a visible {\it sparseness} in the distribution of scintles. This time-frequency occupancy of bright features is likely related to the {\it filling fraction}, $f_{\rm d}$, typically used in the context of scintillation in order to account for the density (or a finite number) of scintles in the dynamic spectrum. In the strong scattering regime, bright features are expected as a natural consequence of an exponential distribution of intensity variations, with separations in time and frequency much larger than the characteristic sizes of scintles \citep{cordes1986}. Observational estimations of scintillation parameters have adopted values in the range $0.2 < f_{\rm d} < 0.5$ \citep{gupta+1994,bhat+1999}, and so are larger than a more conservative value of 0.01 assumed by \citet{cordes1986}. Observations show that this filling fraction tends to vary between pulsars. In order to quantify this time-frequency occupancy of scintles in our data, we introduce the below `sparseness' parameter
\be
s_{\rm d} = { 1 \over  N _{\rm f} \, N _{\rm t}  }  { \sum _{i=1} ^{N _f} 
\sum _{j=1} ^{N _t}  }  s_{\rm on} (f_i, t_j)   
\ee
where
$$
s _{\rm on} (f_{\rm i}, t_ {\rm j}) =
\begin{cases}
	1 & {\rm if\ } S(f_{\rm i}, t_{\rm j}) > 5 \sigma_{\rm off} (f_{\rm i}, t_ {\rm j}) \\
	0 & {\rm otherwise}
\end{cases}
$$

where $\sigma_{\rm off} (f_{\rm i}, t_ {\rm j})$ is the off pulse rms corresponding to the dynamic spectral point $S (f_{\rm i}, t_ {\rm j})$. In other words, this represents the conservative case, where at least one sample within the on pulse window contributing to $S (f_i, t_j)$ exceeds 5$\sigma$ above the mean off pulse level (and thence some detectable pulse energy). In some ways, this can be treated as an upper limit to the filling fraction. 

%

For \psrtwo\ data shown in Fig.~\ref{fig:f6}, values of the \sd\ parameter computed in this manner range from  $\sim$0.05 to $\sim$0.12; the lower values being measured for the highest two frequency bands. It is possible that the relatively low signal-to-noise ratios of these data may be somewhat biasing the estimation of \sd. However, for the data shown in Fig.~\ref{fig:f5}, which are neither limited by signal-to-noise, nor by small-number statistics in terms of the number of scintles, the above estimation still yields $s_d \sim 0.09$.\footnote{For meaningful comparison, the \sd\ parameter is computed on 
$S(f,t)$ that are scaled to match time and frequency resolutions (i.e. 20 s in time and 20 kHz in frequency).} This is comparable to the values for the data shown in Fig.~\ref{fig:f6}, and is significantly lower than that estimated for the dynamic spectra of \psrone\ in Fig.~\ref{fig:f5}, for which we estimate $\sd\sim0.58$. It is possible that such low levels of time-frequency occupancy in terms of scintles is a characteristic of certain lines of sight, where underlying small-scale density structures deviate from the standard Kolmogorov-type distribution. A more quantitative analysis is deferred to a future paper as further observations accrue from our ongoing monitoring project.

\subsection{Chromaticity in Dispersion measure} \label{sec:dm}

The size of the scattering disk is  $ \sim D_{\rm eff} \, \thetadiff $, where  $ \thetadiff \propto \lambda^2 $, thus the sampled ISM volume critically depends on the observing frequency $ f $ (or the wavelength $\lambda $). This means substantially different propagation paths for scattered rays, which may give rise to frequency-dependent (chromatic) DMs. This idea was explored in detail by \citet{cordes+2016}, who caution against the use of low-frequency observations for correcting DM variations. The importance of chromatic DMs thus crucially depends on the degree of scattering, and therefore can be highly  line-of-sight dependent. 

As discussed in \S~\ref{sec:data} (see also Fig.~\ref{fig:f2new}), processing \psrtwo\ data at the catalogued DM results in a striking phase shift of 0.11 turns across our MWA observing band 110$-$220 MHz. It is clear that this cannot be attributed to profile evolution, and implies an excess DM of 0.006 \dmu. This value is several times larger than the peak-to-peak DM variation seen for this pulsar in 6 years of PTA observations \citep{keith+2013}, and may be due to chromatic DM \cite{cordes+2016}.  A similar excess was seen in initial LWA observations of this pulsar \citep{dowell+2013}, and also in ongoing monitoring observations with LWA1, which yields a typical DM of 9.0045 \dmu(K. Stovall, private communication).
Indeed the ISM  sampled by MWA observations would be $\sim$50-100 times larger than that sampled by observations at timing frequencies ($\sim$1-2 GHz), and it is even larger in the LWA observations. The observation of a similar excess in both cases suggests that this is likely a characteristic of the pulsar's line of sight. 

\noindent 

It is, however, possible that there may still be some (small) contribution to the observed DM excess that arises from the frequency evolution of the pulse profile, which is significant for \psrtwo\ (Fig.~\ref{fig:f2}). Assuming temporal stability of the pulse profiles, this would in principle imply a non-variable contribution to $\delta$DM, which can be disentangled if multiple observations are made over long enough time span. While our limited observational data do not allow us to get a handle on that, evidence in support of a persistent excess DM is also hinted at by our multiple observations made with the EDA over a period of $\sim$4 months. A DM excess of similar magnitude is consistently seen in all observations; re-processing MWA data at the correct DM results in nearly aligned pulse profiles across the EDA band, as seen in Fig.~\ref{fig:f3}. A closer examination of EDA observations also suggests that the variable component of the DM (over a 4-month time span) is of the order of $\la$0.001 \dmu; more observational data is needed for a better (and more robust) estimate. 

\noindent

\subsection{Flux densities and Spectra} \label{sec:spec}

Pulsars are generally steep-spectrum radio emitters and the studies based on large samples have shown that the spectral index varies over a large range. The measured values for the mean spectral index \meanalpha vary from $-1.4\pm0.2$ to $-1.8\pm0.2$ \citep[e.g.][]{maron+2000,bates+2013,fabian+2018}. The spectral behaviour at frequencies \la 300 MHz is poorly explored for a large fraction of the known pulsar population, particularly for MSPs, many of which were discovered and studied exclusively at higher frequencies. Low-frequency spectral studies are also complicated by substantial flux variability (by a factor $\sim$3-5) that arises from long-term refractive modulations \citep[e.g.][]{gupta+1994,bhat+1999,wang+2005}. The large frequency span achievable with the MWA and the large fractional bandwidth of the EDA are promising for useful spectral studies, especially for southern MSPs, for which the spectrum at frequencies \la 700 MHz is poorly constrained \citep{dai+2015}. 


\begin{deluxetable}{cccc}[t]
\tablecaption{Spectral index estimates  \label{tab:alpha}} 
\tablecolumns{4}
\tablenum{4}
\tablewidth{0pt}
\tablehead{
\colhead{PSR} &
\colhead{$\alpha _{\rm e}$} &
\colhead{$\alpha _{\rm m}$} &    
\colhead{$\alpha _{\rm p}$} 
}
\startdata
\psrnameone & $-1.1 \pm 0.1 $  &  $-1.41 \pm 0.14 $ &  $ -1.67 \pm 0.04$ \\
\psrnametwo & $-1.73 \pm 0.07$ & $-2.12 \pm 0.07$ & $ -1.96 \pm 0.05$  \\   
\enddata
\tablecomments{$\alpha _{\rm e}$: using EDA data and the measurements from Parkes; $\alpha _{\rm m}$: using MWA data and the measurements from Parkes; $\alpha _{\rm p}$: Parkes measurements (i.e. the mean values of $\alpha _1$ and $\alpha _2$ from \citealt{dai+2015}).}
\end{deluxetable}

\subsubsection{Spectral behavior at low frequencies} 

Using the procedures detailed in \S~\ref{sec:flux}, we estimated the flux density measurements for the data presented in this paper. 
Since our MWA data are limited in both signal-to-noise and temporal resolution (besides multiple caveats of insufficient averaging of flux modulations due to scintillation effects), we limit ourselves to a fairly simple analysis, where we ignore the spectral index variation across the pulse profile, and consider just the spectrum of the average emission (i.e. integrated pulse profiles as a whole). 
Fig.~\ref{fig:f8} presents our measured flux densities for PSRs \psrnameone\ and \psrnametwo, along with the published values from \citet{dai+2015} and \citet{dowell+2013}. Parkes measurements are from regularly sampled PPTA data over a 
6-yr time span, and therefore are more reliable; however,  we note that both pulsars are seen to be highly variable even in the Parkes bands; e.g. rms variability  $\sim$50-70\% is noted for \psrone, whereas \psrtwo\ is seen to vary much more, with rms variation comparable to mean flux densities in the 50 and 10cm data (i.e. 732 MHz and  1.4 GHz). Similarly, large variations are also seen in our MWA observations and therefore the implied spectral behavior can be considered only indicative. EDA measurements are more reliable than those from the MWA, as they are obtained from multiple observations spanning several months; for \psrtwo, nine observations over a 4-month time span,  whereas 6 observations spanning $\sim$8 months were used in the case of \psrone.  Still, this only sparsely samples long-term refractive cycles, and an insufficient averaging of resultant modulations in flux densities can potentially skew our estimates.  

For \psrtwo, excluding LWA measurements,  the combination of EDA and Parkes measurements yields a spectral index estimate, $\alpha _{\rm e} = -1.73 \pm 0.07$, however, a larger value of $\alpha_{\rm m} = -2.12 \pm 0.07$ if we use MWA measurements instead. The reported value of $\alpha _{\rm p} = -1.96\pm0.06$ by \citet{dai+2015} is still within this range, and agrees with our estimates at the 2$\sigma$ level, 
which is encouraging, and reaffirms a generally steeper spectrum of this pulsar. 
Furthermore, it appears that our measurements (in particular, those from the EDA) are not 
consistent with a suggested spectral turnover for this pulsar by \citet{dowell+2013}. In fact, their quoted flux densities are based on a limited number of observations, and hence are subject to possible bias from insufficient averaging of long-term refractive modulations, a caveat also applicable for MWA and EDA measurements. Therefore, the possibility of temporal variations (in flux densities) causing the observed inconsistency cannot be precluded. 

For \psrone, a number of factors prevented us from obtaining reliable estimates of its flux densities. It was difficult to accurately calibrate MWA observations shown in Fig.~\ref{fig:f1} for reliable flux scales because a large number of tiles (58 out of 128) were flagged during the calibration procedure.
The measurements shown in Fig.~\ref{fig:f8} are hence from observations shown in Fig.~\ref{fig:f5}(a) (i.e. over 140-170 MHz, broken up into 12$\times$2.56-MHz contiguous bands). Furthermore, this pulsar also shows substantial variability in apparent brightness -- as much as by a factor of $\sim$6 between different observations. Given these, the indicated $\alpha _{\rm e} = -1.1 \pm 0.07$ (from EDA and Parkes data), which is obtained by excluding the observing epochs when the pulsar was significantly brighter than average, can only be treated as an indicative upper limit (i.e. $|\alpha| > 1.1$). The spectral estimate from MWA and Parkes data is somewhat steeper ($\alpha _{\rm m} = -1.41 \pm 0.14$), however still shallower than the value ($\alpha _{\rm p} = -1.67 \pm 0.04$) from \citet{dai+2015} from Parkes only measurements. We further note that the earlier work of \citet{mcconnell+1996}, which reported the first detection of this pulsar at frequencies below 100 MHz, was inconclusive in terms of a possible turnover at low frequencies. Further observations are needed to explore this in detail. 

\section{Discussion and Future work} \label{disc}

Using the improved capabilities of the MWA, and an SKA prototyping station (the EDA) built utilizing MWA technologies, we have carried out observations of two bright southern MSPs at frequencies below 300 MHz. The primary goal was to demonstrate the advantages of large frequency spans at low radio frequencies for science relating to pulsar emission and probing ISM effects. In this paper, we have largely focused on profile evolution and scintillation studies, as well as measurements of flux densities to investigate the nature of spectral behavior at low frequencies. Even though our analysis is based on data from a limited number of observations, it clearly demonstrates the exciting prospects for low-frequency MSP observations with the MWA and EDA facilities. 

The mean pulse profiles of MSPs tend to show quite a remarkable spectral evolution.  Our observations are generally consistent with the extrapolations based on the recent work of \citet{dai+2015} who performed phase-resolved spectral studies of  a number of MSPs being monitored for the Parkes PTA project. Their data span the frequency range from 732 MHz to 3.1 GHz, and our observations thus extend the frequency coverage down to $\sim$100 MHz. For \psrtwo, there is also some evidence of an excess DM, of the order of $0.006\pm0.003$ \dmu, albeit estimated from a single observation with the MWA. This is six times larger than the peak-to-peak DM variation seen for this pulsar from a 6-yr PPTA data span \citep{keith+2013}. It is however consistent with that inferred from low-frequency detections of this pulsar with the LWA \citep{dowell+2013}. Further, our EDA observations (over a time span of $\sim$4 months) confirm that the observed excess is persistent in the direction of this pulsar. It is possible that part of this excess DM is due to profile evolution at low frequencies, which may give rise to a fixed DM offset, assuming profile stability over time.  Disentangling such a fixed DM contribution (caused by profile evolution) from a variable contribution (caused by the ISM) will require observational data spanning long time spans (months to years), sampled fairly regularly (at a weekly or monthly cadence).   

A similar caveat is also relevant for our flux density and spectral index estimates. While we have been able to meaningfully flux calibrate our pulsar detections, considering the large flux modulations that are typical at low frequencies (due to refractive scintillation), reliable flux density estimates require multiple observations spanning many refractive cycles (typically $\sim$days to weeks for low-DM pulsars). Notwithstanding this caveat, our data are in general agreement with steeper spectra for MSPs, and are consistent with the recent work of \citet{dai+2015}. It appears that a spectral turnover, if at all present, may be occurring at frequencies below the MWA band. 
This is encouraging, particularly when considering low-frequency surveys to find MSPs. The ongoing low-frequency pulsar surveys, such as the Green Bank Northern Celestial Cap (GBNCC) survey \citep{stovall+2014} and the LOFAR survey \citep{bassa+2017}, are already finding MSPs at impressive rates,
which is consistent with the lack of evidence for spectral turnover in our observations. 

In addition to mean flux densities, scintillation properties are also expected to show significant modulations at low frequencies \citep[e.g.][]{gupta+1994,bhat+1999}. The degree of modulation is expected to be  large (as much as $\sim$3--5) in the case of low to moderate DM pulsars, which constitute the majority of PTA pulsar samples. Recent work of \citet{levin+2016} using wide-band observations of northern PTA pulsars  confirm this is generally the case, even at frequencies \ga 1 GHz (see also \citealt{wang+2005}). Hence, in order to meaningfully characterize scintillation effects, observations need to be sampled over multiple refractive cycles, the time scales of which are also longer at low frequencies (weeks to months). A strategy of this kind has also been advocated by \citet{lam+2015} in the context of applying more effective DM corrections using contemporaneous observations. 

In summary, there is quite a compelling case to consider routine observations of MSPs at low frequencies, particularly for PTA pulsars, whose DM and scintillation properties and temporal variabilities will be instructive in ascertaining the ISM noise budget in timing measurements. However, currently the MWA's ability for monitoring purpose is limited by VCS recording, whose data rates and transport logistics constrain observations to maximum 1.5 hr once every 2-3 days. However, the large field-of-view (FOV) of the MWA  can be well exploited for monitoring multiple PTA pulsars at once; e.g. at a frequency of 150 MHz (FOV $\sim$ 450 ${\rm deg^2}$), up to $\sim$5-6 pulsars are covered in a single VCS pointing, thereby yielding high observing efficiency compared to all other currently operational telescopes. Such observations are under way, where the current focus is a modest sample of high-priority southern PTA pulsars. 

In the near-term future, substantial improvements are being planned to enhance the pulsar capabilities of the MWA. These include the ability to reprocess the data by performing an inversion of the poly-phase filterbank  operation prior to the VCS recording, to yield voltage time series data at a much higher time resolution (and hence more amenable for phase-coherent de-dispersion). This will yield even higher quality pulse profiles for MSPs.  The EDA pulsar capability can also be leveraged for robust cross-checks toward verifying the coherent dedispersion functionality for VCS data. The commissioning and verification of polarimetric capability is also under way, and will enable phase-resolved spectral and polarimetric studies of MSPs, similar to the recent work of \cite{nsk+15} at low frequencies, and \citet{dai+2015} at higher frequencies. Furthermore, as more observations accrue from our ongoing monitoring campaign, it will also become possible to undertake a more systematic investigation of subtle effects such as chromatic DM and its relevance for PTA pulsars, as well as MSP spectral behaviour at low frequencies. If MSP spectra turn out to be steeper down to low frequencies, it will strengthen the case to consider low-frequency surveys to find MSPs. Such surveys will be particularly sensitive to finding pulsars with low to moderate DMs, which are most suited for PTA experiments. This will especially benefit future PTA efforts planned with the upcoming MeerKAT \citep{meertime+2017}, and eventually with the SKA \citep{ska+2015}. 

The ongoing developments toward further enhancing the capabilities of the MWA will bring further exciting  prospects for MSP science at low frequencies. 
These include a wider-bandwidth correlator and beamformer, which will eliminate the need to sample the large frequency range by sub-dividing the 30.72-MHz recording bandwidth into smaller sub-bands. Moreover, the increased sensitivity will also help extend low-frequency characterization to a large sample of MSPs, an important exercise that will also facilitate a more effective integration of the MWA into global PTA efforts. In the longer-term future, the much higher sensitivity that will be attainable through a 256-tile MWA, and the observing efficiency achievable via a wide-band beamformer capability, will potentially transform the MWA into a premier low-frequency facility for  high-quality science relating to MSPs. \\

\section{Concluding remarks} \label{sec:conc}  

With the functionality to record voltage data and a suite of post-processing pipelines to perform associated calibration and beam-forming operations, the MWA is rapidly emerging as a promising facility for low-frequency pulsar astronomy in the southern hemisphere. The VCS data can be recorded over a maximum bandwidth of 30.72 MHz, either over a contiguous range, or distributed over a larger range by flexibly sub-dividing into smaller units, to enable a simultaneous sampling of multiple frequencies, and thus effectively providing large fractional bandwidths. This is particularly useful for studying spectral evolution of mean pulse profiles, which is quite remarkable for most MSPs. The data can also be used to investigate a range of ISM effects including the recently-theorised chromatic dispersion that arises from multipath propagation in the ISM, and characterising the strength of turbulence and the degree of scattering in the sight lines toward pulsars. This is especially important for PTA pulsars, and may help converge toward suitable targets and devising optimal observing strategies. We have demonstrated the potential in this area using the case studies of two bright southern MSPs (PSRs \psrnameone\ and \psrnametwo) with the MWA and EDA facilities. 

Systematic monitoring of a large sample of MSPs will help ascertain the role of low-frequency observations for PTA efforts. This is particularly important for southern MSPs which will also become the prime targets for future PTA experiments planned with MeerKAT and the SKA. The MWA is strategically located to undertake this important exercise, and its large field of view can be exploited for simultaneous observations of multiple PTA pulsars.

\acknowledgments
\noindent
{\it Acknowledgements:} 
We thank an anonymous referee for several useful comments that helped improve the presentation and clarity of this paper. 
This scientific work makes use of the Murchison Radio-astronomy Observatory, operated by CSIRO. We acknowledge the Wajarri Yamatji people as the traditional owners of the Observatory site. Support for the operation of the MWA is provided by the Australian Government (NCRIS), under a contract to Curtin University administered by Astronomy Australia Limited. We acknowledge the Pawsey Supercomputing Centre, which is supported by the Western Australian and Australian Governments.
Parts of this research were conducted by the Australian Research Council Centre of Excellence for All-sky Astrophysics (CAASTRO), through project number CE110001020.
The authors acknowledge the contribution of an Australian Government Research Training Program Scholarship in supporting this research.
FK acknowledges support from the Swedish Research Council. 

\vspace{5mm}
\facilities{MWA, EDA}  
\software{DSPSR, PSRCHIVE}



\begin{thebibliography}{}
\bibitem[Abbott et al.(2016)]{ligo2016} Abbott, B.~P., Abbott, R., Abbott, T.~D., et al.\ 2016, Phys Rev Letters, 116, 061102 
\bibitem[Abbott et al.(2017)]{ligo2017} Abbott, B.~P., Abbott, R., Abbott, T.~D., et al.\ 2017, Phys Rev Letters, 119, 161101 
\bibitem[Archibald et al.(2014)]{anne+2014} Archibald, A.~M., Kondratiev, V.~I., Hessels, J.~W.~T., \& Stinebring, D.~R.\ 2014, \apjl, 790, L22 
\bibitem[Arzoumanian et al.(2018)]{nanograv+2016} Arzoumanian, Z., Baker, P.~T., Brazier, A., et al.\ 2018, arXiv:1801.02617
\bibitem[Bailes et al.(2018)]{meertime+2017} Bailes, M., Barr, E., Bhat, N.~D.~R., et al.\ 2018, arXiv:1803.07424 
\bibitem[Bassa et al.(2017)]{bassa+2017} Bassa, C.~G., Pleunis, Z., Hessels, J.~W.~T., et al.\ 2017, \apjl, 846, L20 
\bibitem[Bates et al.(2013)]{bates+2013} Bates, S.~D., Lorimer, D.~R., \& Verbiest, J.~P.~W.\ 2013, \mnras, 431, 1352 
\bibitem[Bhat et al.(2004)]{bhat+2004} Bhat, N.~D.~R., Cordes, J.~M., Camilo, F., Nice, D.~J., \& Lorimer, D.~R.\ 2004, \apj, 605, 759 
\bibitem[Bhat et al.(1999)]{bhat+1999} Bhat, N.~D.~R., Rao, A.~P., \& Gupta, Y.\ 1999, \apjs, 121, 483 
\bibitem[Bhat et al.(2014)]{bhat+2014} Bhat, N.~D.~R., Ord, S.~M., Tremblay, S.~E., et al.\ 2014, \apjl, 791, L32 
\bibitem[Bhat et al.(2016)]{bhat+2016} Bhat, N.~D.~R., Ord, S.~M., Tremblay, S.~E., McSweeney, S.~J., \& Tingay, S.~J.\ 2016, \apj, 818, 86 
\bibitem[Coles et al.(2015)]{coles+2015} Coles, W.~A., Kerr, M., Shannon, R.~M., et al.\ 2015, \apj, 808, 113 
\bibitem[Cordes(1978)]{cordes1978} Cordes, J.~M.\ 1978, \apj, 222, 1006 
\bibitem[Cordes(1986)]{cordes1986} Cordes, J.~M.\ 1986, \apj, 311, 183 
\bibitem[Cordes et al.(2006)]{cordes+2006} Cordes, J.~M., Rickett, B.~J., Stinebring, D.~R., \& Coles, W.~A.\ 2006, \apj, 637, 346 
\bibitem[Cordes \& Rickett(1998)]{cr1998} Cordes, J.~M., \& Rickett, B.~J.\ 1998, \apj, 507, 846 
\bibitem[Cordes \& Shannon(2010)]{cs2010} Cordes, J.~M., \& Shannon, R.~M.\ 2010, arXiv:1010.3785 
\bibitem[Cordes et al.(2016)]{cordes+2016} Cordes, J.~M., Shannon, R.~M., \& Stinebring, D.~R.\ 2016, \apj, 817, 16
\bibitem[Cordes et al.(1985)]{cwb85} Cordes, J.~M., Weisberg, J.~M., \& Boriakoff, V.\ 1985, \apj, 288, 221 
\bibitem[Cordes et al.(1986)]{cordes+1986} Cordes, J.~M., Pidwerbetsky, A., \& Lovelace, R.~V.~E.\ 1986, \apj, 310, 737 
\bibitem[Dai et al.(2015)]{dai+2015} Dai, S., Hobbs, G., Manchester, R.~N., et al.\ 2015, \mnras, 449, 3223  
\bibitem[Demorest et al.(2013)]{nanograv} Demorest, P.~B., Ferdman, R.~D., Gonzalez, M.~E., et al.\ 2013, \apj, 762, 94 
\bibitem[Detweiler(1979)]{detweiler1979} Detweiler, S.\ 1979, \apj, 234, 1100 
\bibitem[de Oliveira-Costa et al.(2008)]{gsm+2008} de Oliveira-Costa, A., Tegmark, M., Gaensler, B.~M., et al.\ 2008, \mnras, 388, 247 
\bibitem[Dowell et al.(2013)]{dowell+2013} Dowell, J., Ray, P.~S., Taylor, G.~B., et al.\ 2013, \apjl, 775, L28 
\bibitem[Dyks et al.(2010)]{dyks+2010} Dyks, J., Wright, G.~A.~E., \& Demorest, P.\ 2010, \mnras, 405, 509 
\bibitem[Foster et al.(2015)]{foster+2015} Foster, G., Karastergiou, A., Paulin, R., et al.\ 2015, \mnras, 453, 1489 
\bibitem[Gil \& Krawczyk(1997)]{gk1997} Gil, J., \& Krawczyk, A.\ 1997, \mnras, 285, 561 
\bibitem[Gupta(1995)]{gupta1995} Gupta, Y.\ 1995, \apj, 451, 717 
\bibitem[Gupta et al.(1994)]{gupta+1994} Gupta, Y., Rickett, B.~J., \& Lyne, A.~G.\ 1994, \mnras, 269, 1035 
\bibitem[Haslam et al.(1982)]{haslam+1982} Haslam, C.~G.~T., Salter, C.~J., Stoffel, H., \& Wilson, W.~E.\ 1982, \aaps, 47, 1 
\bibitem[Hemberger \& Stinebring(2008)]{hs2008} Hemberger, D.~A., \& Stinebring, D.~R.\ 2008, \apjl, 674, L37 
\bibitem[Jankowski et al.(2018)]{fabian+2018} Jankowski, F., van Straten, W., Keane, E.~F., et al.\ 2018, \mnras, 473, 4436
\bibitem[Janssen et al.(2015)]{ska+2015} Janssen, G., Hobbs, G., McLaughlin, M., et al.\ 2015, Advancing Astrophysics with the Square Kilometre Array (AASKA14), 37 
\bibitem[Jones et al.(2017)]{jones+2017} Jones, M.~L., McLaughlin, M.~A., Lam, M.~T., et al.\ 2017, \apj, 841, 125 
\bibitem[Keith et al.(2013)]{keith+2013} Keith, M.~J., Coles, W., Shannon, R.~M., et al.\ 2013, \mnras, 429, 2161 
\bibitem[Kerr et al.(2018)]{kerr+2018} Kerr, M., Coles, W.~A., Ward, C.~A., et al.\ 2018, \mnras, 474, 4637 
\bibitem[Komesaroff et al.(1970)]{komesaroff+1970} Komesaroff, M.~M., Morris, D., \& Cooke, D.~J.\ 1970, \aplett, 5, 37 
\bibitem[Kondratiev et al.(2016)]{vlad+2016} Kondratiev, V.~I., Verbiest, J.~P.~W., Hessels, J.~W.~T., et al.\ 2016, \aap, 585, A128 
\bibitem[Kramer et al.(1998)]{kramer+1998} Kramer, M., Xilouris, K.~M., Lorimer, D.~R., et al.\ 1998, \apj, 501, 270 
\bibitem[Kuzmin \& Losovsky(1996)]{kl96} Kuzmin, A.~D., \& Losovsky, B.~Y.\ 1996, \aap, 308, 91 
\bibitem[Liu et al.(2014)]{liu+2014} Liu, K., Desvignes, G., Cognard, I., et al.\ 2014, \mnras, 443, 375
\bibitem[Lam et al.(2015)]{lam+2015} Lam, M.~T., Cordes, J.~M., Chatterjee, S., \& Dolch, T.\ 2015, \apj, 801, 130 
\bibitem[Lam et al.(2016)]{lam+2016} Lam, M.~T., Cordes, J.~M., Chatterjee, S., et al.\ 2016, \apj, 819, 155 
\bibitem[Lee et al.(2014)]{lee+2014} Lee, K.~J., Bassa, C.~G., Janssen, G.~H., et al.\ 2014, \mnras, 441, 2831 
\bibitem[Lentati et al.(2017)]{lentati+2017} Lentati, L., Kerr, M., Dai, S., et al.\ 2017, \mnras, 466, 3706 
\bibitem[Lentati et al.(2015)]{epta+2015} Lentati, L., Taylor, S.~R., Mingarelli, C.~M.~F., et al.\ 2015, \mnras, 453, 2576 
\bibitem[Levin et al.(2016)]{levin+2016} Levin, L., McLaughlin, M.~A., Jones, G., et al.\ 2016, \apj, 818, 166 
\bibitem[Lorimer \& Kramer(2004)]{handbook} Lorimer, D.~R., \& Kramer, M.\ 2004, Handbook of pulsar astronomy, Cambridge University Press
\bibitem[Maron et al.(2000)]{maron+2000} Maron, O., Kijak, J., Kramer, M., \& Wielebinski, R.\ 2000, \aaps, 147, 195 
\bibitem[Manchester et al.(2013)]{ppta} Manchester, R.~N., Hobbs, G., Bailes, M., et al.\ 2013, PASA, 30, 17
\bibitem[McConnell et al.(1996)]{mcconnell+1996} McConnell, D., Ables, J.~G., Bailes, M., \& Erickson, W.~C.\ 1996, \mnras, 280, 331 
\bibitem[McSweeney et al.(2017)]{mcsweeney+2017} McSweeney, S.~J., Bhat, N.~D.~R., Tremblay, S.~E., Deshpande, A.~A., \& Ord, S.~M.\ 2017, \apj, 836, 224
\bibitem[Meyers et al.(2017)]{meyers+2017} Meyers, B.~W., Tremblay, S.~E., Bhat, N.~D.~R., et al.\ 2017, \apj, 851, 20 
\bibitem[Mitchell et al.(2008)]{mitchell+2008} Mitchell, D.~A., Greenhill, L.~J., Wayth, R.~B., et al.\ 2008, IEEE Journal of Selected Topics in Signal Processing, 2, 707 
\bibitem[Navarro et al.(1997)]{navarro+1997} Navarro, J., Manchester, R.~N., Sandhu, J.~S., Kulkarni, S.~R., \& Bailes, M.\ 1997, \apj, 486, 1019 
\bibitem[Noutsos et al.(2015)]{nsk+15} {Noutsos}, A., {Sobey}, C., {Kondratiev}, V.~I. et al.\ 2015, \aap, 576, A62
\bibitem[Ord et al.(2015)]{ord+2015} Ord, S.~M., Crosse, B., Emrich, D., et al.\ 2015, PASA, 32, e006 
\bibitem[Os{\l}owski et~al.(2011)]{ovh+11} {Os{\l}owski}, S., {van Straten}, W., {Hobbs}, G.~B., {Bailes}, M., \& {Demorest}, P. 2011, \mnras, 418, 1258
\bibitem[Pennucci et al.(2014)]{pennucci+2014} Pennucci, T.~T., Demorest, P.~B., \& Ransom, S.~M.\ 2014, \apj, 790, 93 
\bibitem[Phillips \& Clegg(1992)]{pc92} Phillips, J.~A., \& Clegg, A.~W.\ 1992, \nat, 360, 137 
\bibitem[Prabu et al.(2015)]{prabu+2015} Prabu, T., Srivani, K.~S., Roshi, D.~A., et al.\ 2015, Experimental Astronomy, 39, 73 
\bibitem[Putney \& Stinebring(2006)]{ps2006} Putney, M.~L., \& Stinebring, D.~R.\ 2006, Chinese Journal of Astronomy and Astrophysics Supplement, 6, 233  
\bibitem[Reardon et al.(2016)]{reardon+2016} Reardon, D.~J., Hobbs, G., Coles, W., et al.\ 2016, \mnras, 455, 1751
\bibitem[Sazhin(1978)]{sazhin1978} Sazhin, M.~V.\ 1978, \sovast, 22, 36 
\bibitem[Shannon et al.(2013)]{shannon+2013} Shannon, R.~M., Ravi, V., Coles, W.~A., et al.\ 2013, Science, 342, 334 
\bibitem[Shannon et al.(2014)]{shannon+2014} Shannon, R.~M., Os{\l}owski, S., Dai, S., et al.\ 2014, \mnras, 443, 1463 
\bibitem[Shannon et al.(2015)]{ppta+2015} Shannon, R.~M., Ravi, V., Lentati, L.~T., et al.\ 2015, Science, 349, 1522
\bibitem[Sokolowski et al.(2015)]{marcin+2015} Sokolowski, M., Tremblay, S.~E., Wayth, R.~B., et al.\ 2015, \pasa, 32, e004
\bibitem[Shannon \& Cordes(2017)]{shannon+2017} Shannon, R.~M., \& Cordes, J.~M.\ 2017, \mnras, 464, 2075 
\bibitem[Stinebring et al.(2001)]{stine+2001} Stinebring, D.~R., McLaughlin, M.~A., Cordes, J.~M., et al.\ 2001, \apjl, 549, L97
\bibitem[Stovall et al.(2014)]{stovall+2014} Stovall, K., Lynch, R.~S., Ransom, S.~M., et al.\ 2014, \apj, 791, 67 
\bibitem[Sutinjo et al.(2015)]{sutinjo+2015} Sutinjo, A.~T., Colegate, T.~M., Wayth, R.~B., et al.\ 2015, IEEE Transactions on Antennas and Propagation, 63, 5433 
\bibitem[Taylor et al.(2012)]{lwa2012} Taylor, G.~B., Ellingson, S.~W., Kassim, N.~E., et al.\ 2012, Journal of Astronomical Instrumentation , 1, 1250004-284 
\bibitem[Tingay et al.(2013)]{mwa2013} Tingay, S.~J., Goeke, R., Bowman, J.~D., et al.\ 2013, PASA, 30, 7 
\bibitem[Tremblay et al.(2015)]{tremblay+2015} Tremblay, S.~E., Ord, S.~M., Bhat, N.~D.~R., et al.\ 2015, PASA, 32, e005 
\bibitem[van Haarlem et al.(2013)]{lofar2013} van Haarlem, M.~P., Wise, M.~W., Gunst, A.~W., et al.\ 2013, \aap, 556, A2 
\bibitem[van Haasteren et al.(2011)]{epta} van Haasteren, R., Levin, Y., Janssen, G.~H., et al.\ 2011, \mnras, 414, 3117 
\bibitem[van Straten \& Bailes(2011)]{dspsr2011} van Straten, W., \& Bailes, M.\ 2011, PASA, 28, 1 
\bibitem[van Straten(2013)]{vanstraten2013} van Straten, W.\ 2013, \apjs, 204, 13
\bibitem[van Straten et al.(2001)]{willem+2001} van Straten, W., Bailes, M., Britton, M., et al.\ 2001, \nat, 412, 158 
\bibitem[Walker et~al.(2004)]{wmsz04}{Walker}, M.~A., {Melrose}, D.~B., {Stinebring}, D.~R., \& {Zhang}, C.~M. 2004, \mnras, 354, 43
\bibitem[Wang et al.(2005)]{wang+2005} Wang, N., Manchester, R.~N., Johnston, S., et al.\ 2005, \mnras, 358, 270
\bibitem[Wayth et al.(2017)]{wayth+2017} Wayth, R., Sokolowski, M., Booler, T., et al.\ 2017, \pasa, 34, e034 
\bibitem[Xilouris et al.(1998)]{kiriaki+1998} Xilouris, K.~M., Kramer, M., Jessner, A., et al.\ 1998, \apj, 501, 286 
\bibitem[Yan et al.(2011)]{yan+2011} Yan, W.~M., Manchester, R.~N., van Straten, W., et al.\ 2011, \mnras, 414, 2087 


\end{thebibliography}
\end{document}